\documentclass[fleqn,usenatbib]{mnras}
\usepackage{newtxtext,newtxmath}
\usepackage[T1]{fontenc}

\usepackage{tabularx}
\usepackage{booktabs}
\usepackage{multirow}
\usepackage[table]{xcolor}
\usepackage{adjustbox}
\usepackage{changepage} 
\usepackage{caption}
\usepackage{multicol} 
\usepackage{geometry} 
\usepackage{longtable}
\usepackage{makecell}
\usepackage{colortbl}
\usepackage{array}
\usepackage{arydshln} 
\usepackage{graphicx}
\usepackage{subcaption}
\usepackage{pdflscape}

\arrayrulecolor{black}

\DeclareRobustCommand{\VAN}[3]{#2}
\let\VANthebibliography\thebibliography
\def\thebibliography{\DeclareRobustCommand{\VAN}[3]{##3}\VANthebibliography}

\usepackage{graphicx}
\usepackage{amsmath}
\usepackage{textcomp}
\usepackage{siunitx}
\DeclareSIUnit\angstrom{\text {Å}}
\usepackage{xcolor}
\usepackage{soul}
\usepackage{gensymb}

\sethlcolor{green}

\title[Optical-radio FRB observations]{Contemporaneous optical-radio observations of a fast radio burst in a close galaxy pair}

\author[K. Y. Hanmer et al.]{K. Y. Hanmer,$^{1,2}$\thanks{E-mail: kira.hanmer@gmail.com}
I. Pastor-Marazuela,$^{3}$
J. Brink,$^{1,4,5,6}$
D. Malesani,$^{7}$
B. W. Stappers,$^{3}$
P. J. Groot,$^{1,4,7}$
\newauthor
A. J. Cooper,$^{8}$
N. Tejos,$^{9}$
D. A. H. Buckley,$^{1,4}$
E. D. Barr,$^{10}$
M. C. Bezuidenhout,$^{11}$
S. Bloemen,$^{7}$
M. Caleb,$^{12}$
\newauthor
L. N. Driessen,$^{12}$
R. Fender,$^{8}$
F.~Jankowski,$^{13}$
M. Kramer,$^{10}$
D. L. A. Pieterse,$^{7}$
K. M. Rajwade,$^{8}$
J. Tian,$^{3}$
\newauthor
P. M. Vreeswijk,$^{7}$
R. Wijnands,$^{14}$
and P. A. Woudt$^{1}$
\\
$^{1}$Department of Astronomy, University of Cape Town, Private Bag X3, Rondebosch 7701, South Africa\\
$^{2}$High Energy Physics, Cosmology \& Astrophysics Theory (HEPCAT) Group, Department of Mathematics \& Applied Mathematics,\\ University of Cape Town, Cape Town, 7700, South Africa\\
$^{3}$Jodrell Bank Centre for Astrophysics, University of Manchester, Oxford Road, Manchester M13 9PL, UK\\
$^{4}$South African Astronomical Observatory, P.O. Box 9, Observatory, 7935, South Africa\\
$^{5}$Leibniz-Institut für Astrophysik Potsdam (AIP), An der Sternwarte 16, 14482 Potsdam, Germany\\
$^{6}$Institute for Physics and Astronomy, University of Potsdam, Karl-Liebknecht-Str. 24/25, 14476 Potsdam, Germany\\
$^{7}$Department of Astrophysics/IMAPP, Radboud University, P.O. Box 9010, 6500 GL, Nijmegen, The Netherlands\\
$^{8}$Astrophysics, The University of Oxford, Denys Wilkinson Building, Keble Road, Oxford OX1 3RH, UK\\
$^{9}$Instituto de F\'{ı}sica, Pontificia Universidad Cat\'{o}lica de Valpara\'{ı}so, Casilla 4059, Valpara\'{ı}so, Chile\\
$^{10}$Max-Planck-Institut für Radioastronomie, Auf dem Hügel 69, D-53121 Bonn, Germany\\
$^{11}$Centre for Space Research, North-West University, Potchefstroom 2531, South Africa; Department of Mathematical Sciences, University of South Africa,\\ Cnr Christiaan de Wet Rd and Pioneer Avenue, Florida Park, 1709, Roodepoort, South Africa\\
$^{12}$Sydney Institute for Astronomy, School of Physics, The University of Sydney, New South Wales 2006, Australia\\
$^{13}$ LPC2E, OSUC, Univ Orleans, CNRS, CNES, Observatoire de Paris, F-45071 Orleans, France\\
$^{14}$ Anton Pannekoek Institute for Astronomy, University of Amsterdam, P.O. Box 94249, 1090 GE Amsterdam, The Netherlands\\
}

\date{Accepted XXX. Received YYY; in original form ZZZ}

\pubyear{2024}

\begin{document}
\label{firstpage}
\pagerange{\pageref{firstpage}--\pageref{lastpage}}
\maketitle

\begin{abstract}
We present the MeerKAT discovery and  MeerLICHT contemporaneous optical observations of the Fast Radio Burst (FRB) 20230808F, which was found to have a dispersion measure of $\mathrm{DM}=653.2\pm0.4\mathrm{\,pc\,cm^{-3}}$. FRB\,20230808F has a scattering timescale $\tau_{s}=3.1\pm0.1\,\mathrm{ms}$ at $1563.6$\,MHz, a rotation measure $\mathrm{RM}=169.4\pm0.2\,\mathrm{rad\,m^{-2}}$, and a radio fluence $F_{\mathrm{radio}}=1.72\pm0.01\,\mathrm{Jy\,ms}$. We find no optical counterpart in the time immediately after the FRB, nor in the three months after the FRB during which we continued to monitor the field of the FRB. We set an optical upper flux limit in MeerLICHT's $q$-band of $11.7\,\mathrm{\micro Jy}$ for a 60\,s exposure which started $\sim3.4$\,s after the burst, which corresponds to an optical fluence, $F_{\mathrm{opt}}$, of $0.039\,\mathrm{Jy\,ms}$ on a timescale of $\sim3.4$\,s. We obtain an estimate for the $q-$band luminosity limit of $vL_{v}\sim 1.3\times10^{43}\,\mathrm{erg\,s^{-1}}$. We localise the burst to a close galaxy pair at a redshift of $z_{\mathrm{spec}}=0.3472\pm0.0002$. Our time delay of $\sim3.4$\,s between the FRB arrival time and the start of our optical exposure is the shortest ever for an as yet non-repeating FRB, and hence the closest to simultaneous optical follow-up that exists for such an FRB.\
\end{abstract}


\section{Introduction}
Fast Radio Bursts (FRBs), first discovered in 2007, are bright, millisecond-duration pulses of radio emission. FRBs have dispersion measures (DMs) that are far greater than the expected DM contribution of the Milky Way galaxy, meaning that they are of extragalactic origin. The first FRB was found in archival pulsar data from 2001 which made use of Murriyang, the Parkes 64-m radio telescope \citep{Lorimer_2007}, with an additional four bursts discovered in 2013, confirming them as a population \citep{Thornton_2013}.

FRBs are sub-divided into two main groups: repeaters and non-repeaters, although there is ongoing debate as to whether or not these are two fundamentally different populations. Non-repeating FRBs are those which have only been detected once, and out of the more than $700$\footnote{\href{https://www.wis-tns.org/}{https://www.wis-tns.org/}} FRBs discovered and published to date, most of them fall into this category. Repeating FRBs, on the other hand, are those which have been observed to burst multiple times, and $\sim$60 of this type have been published to date \citep{Chime_2023}. The isotropic energies of FRBs span a range from $\sim10^{35}$ to $10^{43}\,\mathrm{erg}$.

The DMs of known FRBs span from $87.82\,\mathrm{pc\,cm^{-3}}$ for the repeater FRB\,20200120E, localised to a globular cluster in M81 \citep{Bhardwaj,Kirsten}, to $3338\,\mathrm{pc\,cm^{-3}}$, for FRB\,19920913A \citep{Crawford_2022}. The most distant localised FRB is FRB\,20220610A, which has a redshift of $z=1.016\pm0.002$ \citep{Ryder}. Recently, however, \cite{Connor_2024} published the newly discovered FRB\,20230521B, which they report as having a redshift of $z=1.354$. This measurement is tentative for the time being, as it is derived from a single line presumed to correspond to $\mathrm{H\alpha}$. If confirmed, it would exceed the redshift of FRB\,20220610A.

The origin of FRBs remains unknown, although numerous models have been suggested. Some of these models propose that FRBs originate from compact objects, and there are numerous ones which predict that at least some FRBs come from magnetars (e.g. \citealt{Metzger_2017, Beloborodov_2017, Beloborodov_2020, Kumar_2017, Yang_2018}). Cataclysmic models were at first suggested as potential explanations for FRBs (e.g. \citealt{Falcke_2014}), but these were ruled out as an explanation for all FRBs once some were found to repeat (\citealt{Spitler_2016}; see \citealt{Platts_2019} and \citealt{Zhang_2023} for a comprehensive review of FRB progenitor models). The detection of a bright radio burst \citep{2020Natur.587...54C,bochenek_2020} from SGR 1935+2154 \citep{Cummings_2014, Lien_2014, Israel_2016}, a Galactic magnetar, gave some evidence for the idea that magnetars can cause some FRBs. The radio burst from SGR 1935+2154 is sometimes called the first "Galactic" FRB (FRB\,20200428), despite being significantly less luminous than FRBs. There are two main classes of models associated with magnetar progenitors: magnetospheric models, in which the emission comes from close to the surface of the neutron star \citep{Katz_2016,Kumar_2017,Wadiasingh_2019,Kumar_2020,Lyutikov_2020,Cooper_2021}, and maser shock models, in which the pulse results from maser emission in magnetised shocks \citep{Beloborodov_2017,Metzger_2019,Beloborodov_2020}.

It remains unclear whether or not non-repeaters and repeaters have different underlying mechanisms, or whether the non-repeaters are simply repeating at much lower rates, and we therefore have not yet observed their repetition. However, there are some differences which have been observed in the burst structures of non-repeaters versus repeater FRBs: repeaters often display what is known as the "sad trombone effect" \citep{Hessels_2019, CHIME_2019, Fonseca_2020, Day_2020}, whereby bursts have multiple components that drift downwards in frequency. Non-repeaters, on the other hand, can have bursts with a single component that can be either broad or narrowband, or bursts with multiple components which all peak at the same frequency. Additionally, repeaters tend to be narrower in frequency and wider in time than non-repeaters \citep{Pleunis_2021}.

FRBs have not yet been conclusively observed at other wavelengths, and as such, no counterpart of statistical significance has been found, although there have been multiple attempts (e.g. \citealt{Hardy_2017,Aartsen_2018,Acciari,Callister_2016,Eftekhari,Petroff,Zhang,trudu, Hiramatsu_2023, Niino, gamma-ray, kilpatrick_2021, kilpatrick_2023, Pearlman_2023, Nunez_2021}). A confirmed optical counterpart would help to rule out at least some of the currently suggested models for FRB emission mechanisms. For example, the synchrotron maser model of FRBs predicts a multi-wavelength afterglow, while most magnetospheric models do not. The detection of an optical afterglow would provide evidence in favour of the former model. For the burst associated with SGR 1935+2154, \cite{Margalit_2020} interpreted the X-ray counterpart as the X-ray afterglow of the FRB. However, the non-detections of an optical afterglow for FRB\,20200428 (e.g. \citealt{Bailes_2021}) greatly constrain the maser shock model \citep{Cooper_2022}. In this work, we present our discovery of the to date non-repeating FRB\,20230808F, and our search for an optical counterpart, utilising almost simultaneous optical-radio observations.

\section{Radio Observations and data reduction}\label{sec:Radio}
\subsection{MeerTRAP pipeline and FRB\,20230808F detection}\label{sec:Meertrap}
MeerKAT is a radio telescope array consisting of 64 13.5-m dishes, located in the Karoo in South Africa \citep{Jonas_2016}. MeerKAT's detection of FRBs is carried out by the Transient User Supplied Equipment (TUSE) instrument of MeerTRAP (Meer(more) TRAnsients and Pulsars; \citealt{sanidas_2018,Rajwade_2022,Bezuidenhout_2022,Jankowski_2022}), which conducts searches for single pulses concurrently in both coherent and incoherent beams. MeerTRAP is a commensal survey with MeerKAT, and observes simultaneously with all MeerKAT Large Survey Projects and some other projects. The coherent beams (CBs) are formed by a beam-forming instrument called the Filterbanking Beamformer User Supplied Equipment (FBFUSE; \citealt{Chen_2021,Barr_2017}), which coherently combines the signals from the dishes of the telescope array. FBFUSE and TUSE together form the MeerTRAP backend, which can form up to 768 CBs that are highly sensitive and have a total field of view (FoV) of $\sim$0.4\,$\mathrm{deg^2}$. This is only a fraction of the primary FoV of MeerKAT, which is $\sim$1\,$\mathrm{deg^2}$ at $1284$\,MHz, although this fraction also depends on the elevation of the source being observed by the telescope \citep{Chen_2021,Rajwade_2022}. The incoherent beam (IB) is formed by summing the signals from all of the available antennas, resulting in a less sensitive beam which samples the entire $\sim$1\,$\mathrm{deg^2}$ FoV and which usually has around one fifth of the sensitivity of the CBs. AstroAccelerate\footnote{\href{https://github.com/AstroAccelerateOrg/astro-accelerate}{https://github.com/AstroAccelerateOrg/astro-accelerate}} \citep{dimoudi2015pulsar,Dimoudi_2018,adámek2016realtime,adámek2017improved,adámek2018gpu}, a GPU-based single pulse search pipeline, is used to perform a real-time search for dispersed bursts, by incoherently de-dispersing the data in the DM range 0--5118.4\,pc\,$\mathrm{cm^{-3}}$ (see \citealt{Caleb_2020} for further details about the search process).

FRB\,20230808F was detected by MeerTRAP at the $L$-band, which is centered on 1284\,MHz and has a bandwidth of 856\,MHz, on 2023 August 8. The arrival time at the top of the observing band for the FRB was 03:49:15.888 UT and its dispersion measure was found to be $\mathrm{DM}=653.2\pm0.4$\,pc\,$\mathrm{cm^{-3}}$. The time of arrival at infinite frequency for the FRB was calculated using the arrival time at the top of MeerKAT's \textit{L}-band (1712\,MHz) and taking into account the frequency dispersion in the burst arrival time, due to the ionised interstellar and intergalactic medium \footnote{The dispersive time delay is calculated using the dispersion constant $D=4.1488\times 10^3\,\mathrm{MHz^{2}\,pc^{-1}\,cm^{3}\,s}$.}. This resulted in an arrival time at the MeerKAT site of 03:49:14.963 UT on 2023 August 8 at infinite frequency.

\subsection{Localisation}
The detection of FRB\,20230808F in a coherent beam and  the incoherent beam provides an approximate region of the sky from which the burst originated. The burst passed the detection and machine learning criteria within 45 seconds after the detection, triggering a transient buffer (TB) data dump, whereby high-resolution data around the time of a transient event is captured and stored. The TB data was used to localise the FRB, as well as to obtain the beam-formed data which has a higher spectro-temporal resolution. A detailed description of the MeerTRAP TB system and FRB localisation can be found in \cite{Rajwade_2024}.

\begin{figure}
\begin{center}
	\includegraphics[width=0.5\textwidth]{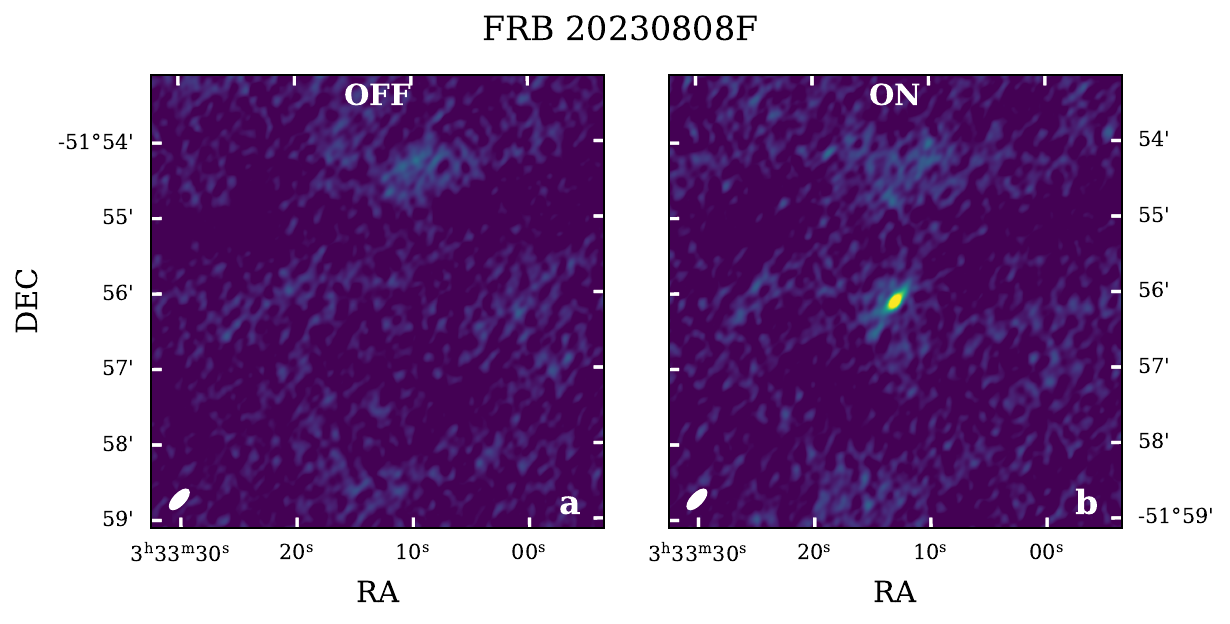}
    \caption{"Off" and "on" cleaned images showing the burst of FRB\,20230808F. Each image has a total exposure time of approximately $14.7$\,ms.}
    \label{fig:on_off}
    \end{center}
\end{figure}

The TB data provides a complex voltage dataset, which contains all of the voltages received by MeerKAT's antennas that were active at the time of the burst detection. The voltage data contain 300\,ms of signal tracing the dispersion curve of the FRB, which is divided into 64 equal sub-bands. The data were imaged using \texttt{WSClean} \citep{wsclean}, producing a separate image for each frequency sub-band. Following this, we used Python to produce the images which would be used to search for the FRB. This entailed visually inspecting the full 300-ms images so as to flag and then discard the sub-bands which contained radio frequency interference (RFI), before the remaining images were averaged to produce an RFI-free image that was used as a reference image to look for the burst. In addition, images with shorter integration times were produced, in order to determine whether any new source appeared during the central milliseconds, which would be the FRB. In the case of FRB\,20230808F, the 300\,ms were divided into 31 time intervals, and these smaller timescales were imaged, again averaging the RFI-free frequencies. Difference images were then produced by subtracting the full integration time image from each of the smaller time interval images. A source was found in the difference images for the central time intervals, and, with a location coincident with the CB where it was detected, we concluded that this was the FRB. The full 300\,ms were then divided into 489 time intervals in order to enable selection of all images in which the burst was detected, from which we produced an "on" image. Similarly, an "off" image was created containing the same number of time bins, but where the FRB was not present. These on and off images, as well as a new full integration time image, were produced and cleaned with \texttt{WSClean}\footnote{
WSClean is an imaging algorithm for radio data that uses w-stacking to correct for w-term effects after the FFT. Further details can be found at: \href{ https://wsclean.readthedocs.io/en/latest/ }{https://wsclean.readthedocs.io/en/latest/}}, using the following parameters: the stopping criteria were 100 iterations, or a threshold of 0.01 (arbitrary units). We applied a \texttt{Cotton-Schwab} \citep{Schwab_1984} cleaning with major iteration gain of 0.8, and auto-masking threshold of $\sigma=3$. We used a Briggs weighting with a robustness parameter of $-0.3$ with a weight rank filter of 3. Lastly, we applied a W-gridding mode. The final images were centred at the phase centre of the observation, they have a pixel size of $1 \, \text{arcsec}^2$, and a size of $4096\times4096$ pixels. The clean on and off images, zoomed at the location of the FRB, are shown in Figure~\ref{fig:on_off}.

Following this, astrometry was performed using the method described in \cite{driessen2023frb}, by running the Python Blob Detector and Source Finder (\texttt{pyBDSF}; \citealt{pybdsf}) algorithm to identify the positions of the sources in the images. The astrometric transformation was obtained by searching other radio catalogues with known sources in the FoV, which were used as references. Although we prioritise using catalogues with a positional accuracy <0.5”, the FoV of FRB\,20230808F is not covered by any such catalogues containing sufficient sources to perform the astrometric correction ($\gtrsim5$). While the Rapid ASKAP Continuum Survey (RACS-mid; \citealt{Duchesne_2023}) contains sufficient sources within the FRB FoV, its sources have systematic positional uncertainties of 1–2”. Thus, as described in \cite{Rajwade_2024}, we use the Radio Fundamental Catalog (RFC) \footnote{RFC: \href{ http://astrogeo.org/rfc/ }{http://astrogeo.org/rfc/}} to correct the RACS-mid source positions. The RFC has source positions with milli-arcsecond accuracy, and while it does not contain enough sources to perform the astrometric correction of the FRB FoV, it is enough to correct the RACS-mid source positions within a 5 degree radius. We then used the RACS-mid corrected source positions to correct the MeerKAT images. The astrometric correction was performed with the python package   \texttt{Astroalign} \citep{astroalign}, and only sources classified as point-like were used. The transformation was applied to the on, off, and full integration time images, and the corrected FRB coordinates were then obtained from the transformed on image.  The position errors for the FRB were obtained by summing in quadrature the position error from \texttt{pyBDSF}, and the astrometric error from the two transformations, which is defined as the mean angular separation between the sources in the reference catalogue and the corrected coordinates of sources in the target catalogue. In our case, the reference and target catalogues for the first astrometric correction were RFC and RACS, respectively, and for the second correction they were RACS and MeerTRAP.  FRB\,20230808F was localised to a position of $\mathrm{RA\,(J2000)=03^{h}33^{m}12^{s}.99\pm0^{s}.03}$ and $\mathrm{Dec\,(J2000)=-51\degree56'07\farcs02}\pm0\farcs50$.

\subsection{Burst structure and timing analysis}
\begin{figure}
\begin{center}
	\includegraphics[width=0.4\textwidth]{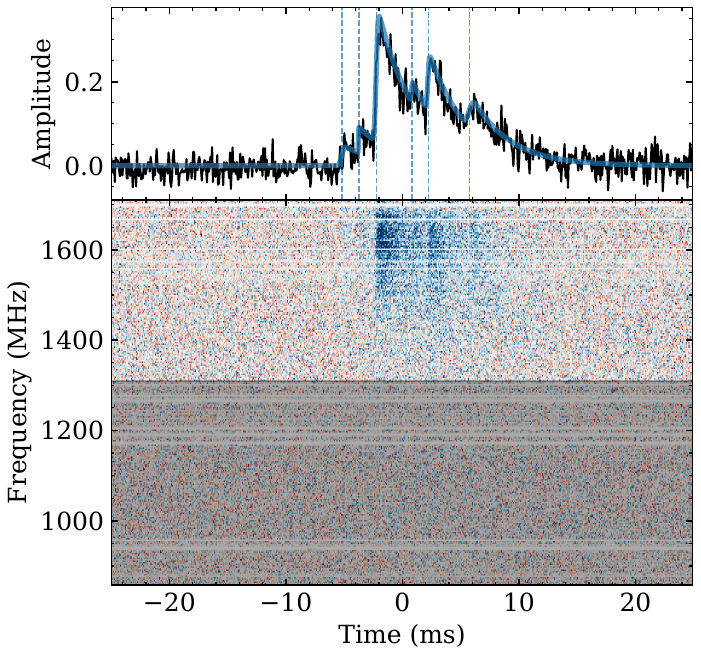}
    \caption{The dynamic spectrum and burst profile of FRB\,20230808F. The black solid line in the top panel shows the averaged pulse profile, the solid blue line shows the fit to a multi-component scattered Gaussian, and the dashed blue lines indicate the positions of the six burst components. The dynamic spectrum is shown in the bottom panel, with the shaded grey area showing the region outside of the FWTM of the spectrum.}
    \label{fig:Profile}
    \end{center}
\end{figure}

Using the localised FRB coordinates, the TB data were beamformed at the correct FRB position, resulting in a boost in the signal-to-noise ratio  (S/N; see \citealt{Rajwade_2024} for further details of this process). The TB data had a higher spectro-temporal resolution, and were coherently de-dispersed at the DM of the burst, eliminating intra-channel smearing. The beamformed data were used to resolve the FRB components, along with the scattering timescale. To fit the temporal structure of the FRB, the frequency range where the burst emission was most visible - primarily at the upper end of the observing bandwidth - was selected to enhance the S/N of the pulse profile. This involved averaging the dynamic spectrum in time, fitting it with a Gaussian function, and then selecting frequencies within the Full Width at Tenth Maximum (FWTM) of the fitted Gaussian. For FRB 20230808F, this range spanned from 1415.25 MHz to 1738.23 MHz. The upper frequency limit is above the maximum of MeerKAT's $L$-band, so all channels above 1415.25\,MHz up to 1712\,MHz were included. In order to fit the pulse to a multi-component model with scattering, a semi-automated fitting routine was created to select the locations of the components, fit multi-component scattered Gaussians, and then find the optimal number of components and scattering timescale. The fitting routine models the pulse profile of an FRB using a multi-component scattered Gaussian model. This is done by first smoothing the pulse profile with a Blackman window of length 15 and identifying local maxima within the smoothed pulse profile. A range for the number of components (minimum and maximum) is provided as input, and the largest peaks in the smoothed profile are used as initial guesses for the component locations. The profile is then fitted iteratively using \texttt{lmfit.Minimizer} and custom functions that convolve multiple gaussians with a one-sided exponential decay to add scattering. The fitting process minimizes the difference between the model and the data, adding components iteratively by smoothing residuals and identifying additional peaks, until the maximum component number is reached. The optimal number of components is then chosen based on the lowest Bayesian Information Criterion (BIC), yielding the number of components, scattering timescale, and the arrival time, amplitude, and width of each component. This resulted in an optimal number of components of 6, along with the time of arrival (ToA) of each component, and a scattering timescale $\tau_{s}=3.1\pm0.1$\,ms. The latter is measured at 1563.6\,MHz, which is the central frequency of the burst bandwidth. The locations of the 6 burst components are shown in Figure~\ref{fig:Profile}.

The method described in \cite{Pastor_Marazuela_2023} was used to test for periodicity in the burst structure and to obtain the component arrangement. The ToAs as a function of component number were fitted to a linear function of the form:
\begin{equation}\label{Eq:lmfit}
    t_{i}=\Bar{d}n_{i}+T_{0},
\end{equation}
where $t_{i}$ is the $i$-th arrival time, $\Bar{d}$ is the mean spacing between the sub-components of the burst, $n_i$ is the $i$-th component number, and $T_0$ is the first ToA. The fitting was conducted using the Python package \texttt{lmfit}\footnote{\href{https://pypi.org/project/lmfit/}{https://pypi.org/project/lmfit/}} \citep{matt_newville_2024_12785036}. A least squares minimisation technique, weighted by the inverse of the ToA errors, was used to determine the mean sub-component separation $P_\mathrm{{sc}}$, equivalent to $\Bar{d}$, and the goodness of fit was assessed using the reduced chi squared statistic, $\chi_{r}^2$. This yielded $P_\mathrm{{sc}}=1.55\pm0.03$\,ms.

To test the significance of the  periodicity, $10^5$ bursts with six sub-components and random separations were simulated and the same linear fit described by Eq \ref{Eq:lmfit} was performed, with the resulting distribution of $\chi_{r}^2$ values compared to the $\chi_{r}^2$ for the FRB. The sub-burst separations were simulated using two different distributions, namely a rectangular and a Poissonian distribution, and using different exclusion parameters. The exclusion parameter $\eta$ is the minimal separation fraction that we consider would be resolved by the fitting routine. Using rectangular simulations, the periodicity significance was equal to $2.22\,\mathrm{\sigma}$, while for the Poissonian distribution, the significance was $2.11\,\mathrm{\sigma}$. In both cases, therefore, we do not find that the burst is significantly periodic.

\subsection{Polarisation and rotation measure}
\begin{figure}
\begin{center}
	\includegraphics[width=0.3\textwidth]{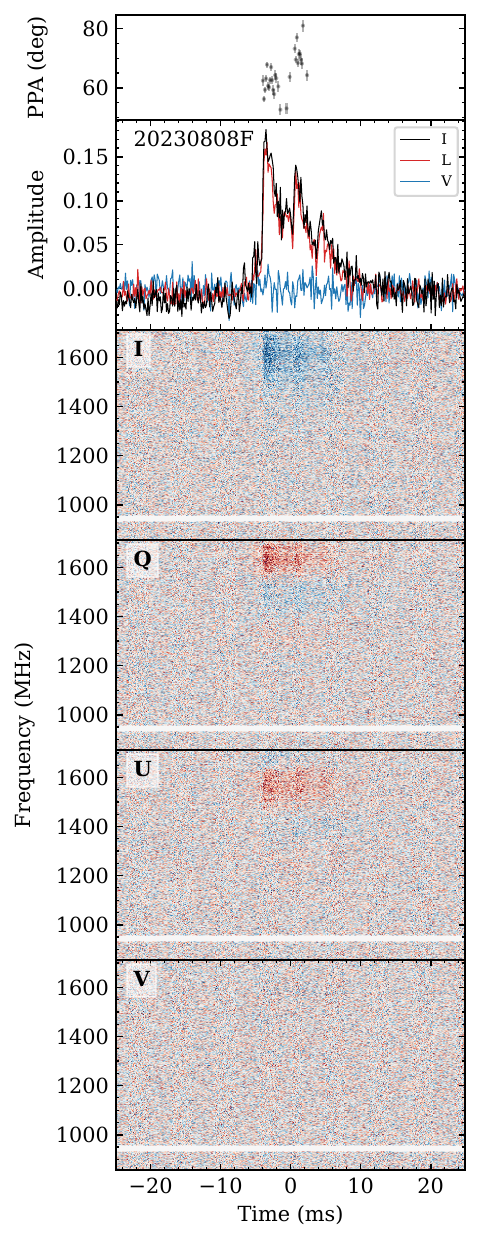}
    \caption{The Stokes I, Q, U and V data for FRB\,20230808F are shown in the four bottom panels, prior to the de-rotation of Stokes Q/U. The top two panels show the PPA and pulse profile, respectively, both after de-rotating Q/U, and a scattering tail is visible in the pulse profile. The blue in the Stokes panels indicates positive intensity values, while red is associated with negative values.}
    \label{fig:Stokes}
    \end{center}
\end{figure}

The TB data provide full Stokes I, Q, U and V information, which was used to obtain the rotation measure and intrinsic polarisation properties of the burst. The Faraday rotation measure (RM) for FRB\,20230808F was measured using the \texttt{RM-Tools} algorithm \citep{RM_tools}, which applies the RM synthesis technique \citep{Burn_1966,Brentjens_2005}. This yielded a $\mathrm{RM}=169.4\pm0.2$ rad $\mathrm{m}^{-2}$. We then de-rotated the Stokes Q/U data at the measured RM, which removed any variations with frequency and enabled us to study the intrinsic emission. The average linear polarisation factor, L/I, was found to be $106.6\pm14.8$ per cent, while the average circular polarisation factor, V/I, was found to be $10.4\pm7.8$ per cent. These values are consistent with $100$ per cent linear and $0$ per cent circular polarisation fractions, which is what is commonly observed in repeating and some one-off FRBs (e.g. \citealt{Cho_2020, Day_2020, Luo_2020_polarisation, Michilli_2018}). The polarisation position angle (PPA) is given by:
\begin{equation}
    \psi(t)=\frac{1}{2}\tan^{-1}\frac{U_{\mathrm{derot}}(t)}{Q_{\mathrm{derot}}(t)}.
\end{equation}
The PPA is shown in the top panel of Figure~\ref{fig:Stokes}, and we found that it appears to step from $60\degree$ to $80\degree$ over the duration of the burst, which is consistent with behaviour seen in multiple pulsars and magnetars, and other FRBs (e.g. \citealt{Cho_2020, Luo_2020_polarisation}).

\section{Host galaxy and optical counterpart search}

\subsection{Identification of host galaxy}\label{sec:host_gal}
\begin{figure}
	\includegraphics[width=\columnwidth]{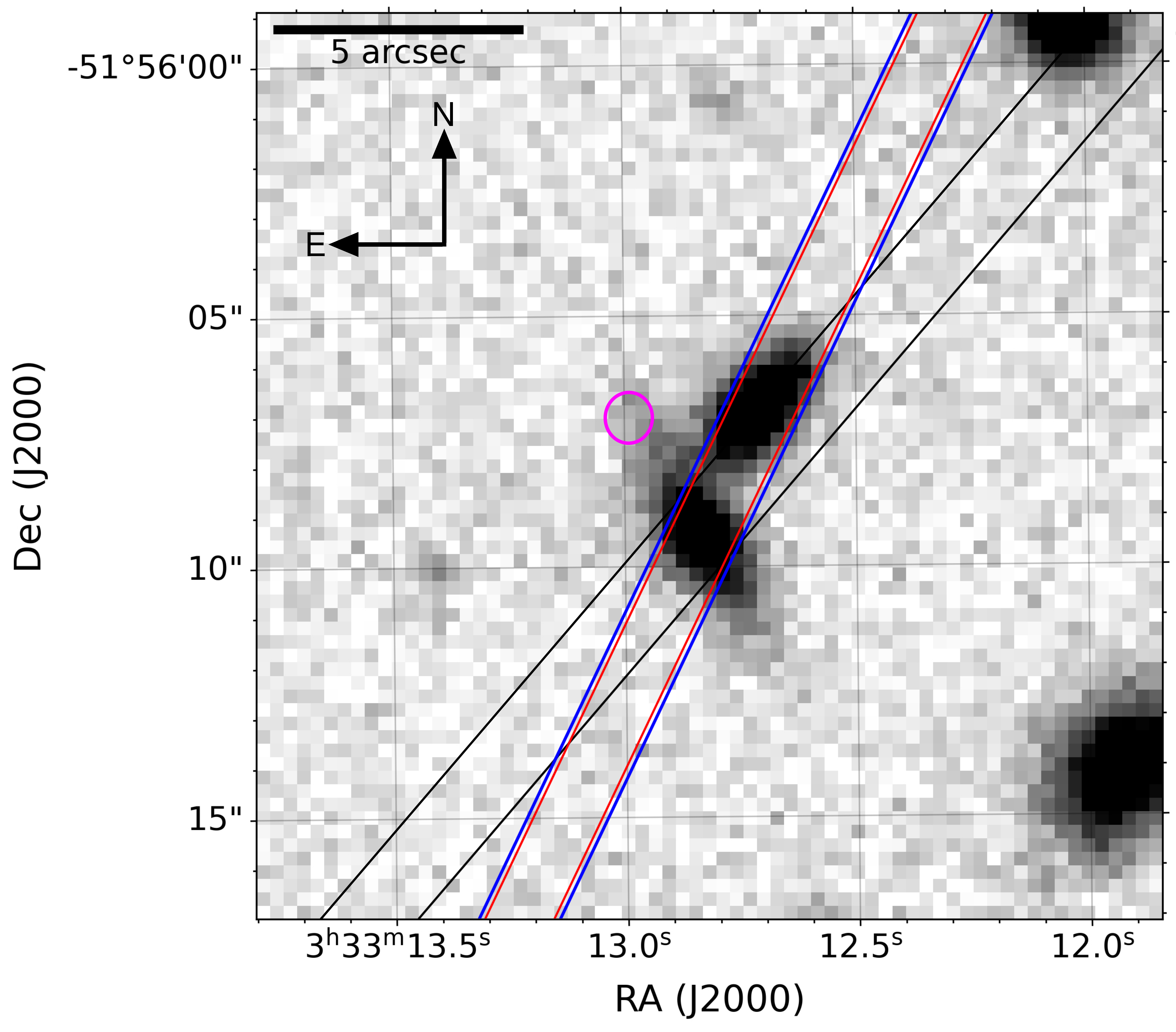}
    \caption{The two host galaxy candidates can be seen in the $i$-band image above, taken from the Dark Energy Survey's (DES; \protect\citealt{thedarkenergysurveycollaboration2005dark}) second public data release (hereafter DR2; \protect\citealt{DES_dr2}). The magenta ellipse indicates the position with uncertainties of FRB\,20230808F. The black parallel lines indicate the slit position for the November SALT observations, and the blue and red parallel lines show the slit for the 2023 December 11 and 2023 December 16 observations, respectively.}
    \label{fig:DES}
\end{figure}

After localising the FRB, we searched at the coordinates of the burst in the Dark Energy Spectroscopic Instrument (DESI; \citealt{desi_2016}) Legacy Survey's tenth data release (hereafter DESI DR10), and it was apparent that there were two potential host galaxies, although the burst location appeared closer to one of the two. From visual inspection, the two galaxies appeared to possibly be interacting, as can be seen, along with the FRB coordinates, in Figure~\ref{fig:DES}. The coordinates of the two galaxies are $\mathrm{RA\,(J2000)=03^{h}33^{m}12^{s}.70,\,Dec\,(J2000)=-51\degree56'06\farcs72}$ for the northernmost galaxy (hereafter north galaxy) and $\mathrm{RA\,(J2000)=03^{h}33^{m}12^{s}.83,\,Dec\,(J2000)=-51\degree56'09\farcs27}$ for the southernmost galaxy (hereafter south galaxy).

We run the Probabilistic Association of Transients to their Hosts (PATH) algorithm \citep{path} for a sample of galaxies from DESI DR10 within a FoV of $30$\,arcseconds around the FRB position. PATH is a Bayesian algorithm that provides posterior probabilities ($P(O_i|x)$) for host associations based on observables and priors that depend only on photometry and astrometry (including the FRB position uncertainty). We used photometry in the $r$-band with an exponential profile prior centred in each candidate galaxy with a characteristic scale given by their half-light radii ($r_e$) up to a maximum of $10\times r_e$. We also used a prior on the true host not being included in the catalog (i.e. ``unseen'') of $P(U)=0.05$. The rest of the parameters were set as default in the PATH repository.\footnote{\href{https://github.com/FRBs/astropath}{https://github.com/FRBs/astropath}} We obtained $P(O_i|x)=0.59$ and $P(O_i|x)=0.37$, for the north and south galaxy respectively. Changing the characteristic scale of the prior exponential profile to $r_e/2$ as suggested by \citet{shannon2024} does not significantly change the posteriors. This analysis provides support to the association of the FRB with either of the two closest identified galaxies, with
the sum of these posterior probabilities being larger than $0.9$. We note that, while PATH assigns a higher association probability to the north galaxy, the algorithm does not take into account the ellipticity and orientation of the host galaxy candidates. From the optical image, shown in Figure~\ref{fig:DES}, the FRB localisation seems to fall within the extension of the south galaxy.

\cite{Zou_2022} calculated the photometric redshifts for around 293 million galaxies with \textit{i<24} in DES DR2, including the two potential host galaxies. The photometric redshift for the north galaxy from \cite{Zou_2022} is $z_{\mathrm{phot}}=0.42\pm0.06$, while that of the south galaxy is $z_{\mathrm{phot}}=0.53\pm0.07$. These two redshifts agree within uncertainties, implying that the galaxies may be a close, potentially interacting pair. The DESI DR10 magnitudes for the galaxies are presented in Table \ref{tab:DESI_mags}, and various properties of the galaxies resulting from the SED fitting are shown in Table \ref{tab:SED_prop}. The star formation rates (SFRs) obtained for both galaxies are negative, which likely means that the true SFR is low, consistent with zero. We also note the comment made by the authors, which cautions the reader against using the star formation rate and stellar age parameters from their catalogue.

\subsection{MeerLICHT optical counterpart search}
MeerLICHT, situated at the South African Astronomical Observatory site near Sutherland in South Africa, is a fully robotic, wide-field optical telescope. The telescope has a 0.65 m primary mirror and a 2.7 $\mathrm{deg^2}$ field of view, and was built as a prototype for the BlackGEM array in Chile \citep{Groot_2024}. MeerLICHT has a filter wheel which consists of the Sloan Digital Sky Survey (SDSS; \citealt{SDSS}) \textit{u, g, r, i} and \textit{z} filters, and a wider \textit{q} band which spans the range of 440--720 nm \citep{Groot_2024}. MeerLICHT has a declination upper limit of $+30$\textdegree, and pointing limits of $>20$\textdegree above the horizon in elevation and hour angle limit greater than $-3$ hours and less than 4.5 hours. When the aforementioned conditions are met, MeerLICHT co-observes with MeerKAT, and if syncing cannot be established for one or more reasons, the telescope either follows its backup program of observing regions of Local Universe mass over-densities, follows up other high priority targets, or it builds up reference images for its survey of the southern sky.

When MeerLICHT is synced with MeerKAT, it observes in one-minute exposures, and every second image is a $q$-band exposure in the sequence \textit{q, g, q, i, q, u, q, r, q, z, q}. When not synced with MeerKAT, MeerLICHT observes in the filter sequence \textit{q, u, i, q, u, i}, etc., also with an integration time of one minute. Regardless of whether or not it is linked with MeerKAT, there are approximately 25 seconds of overhead time in between each MeerLICHT exposure, which allows for filter change, re-pointing of the telescope, and image readout.

MeerLICHT images are reduced using two pieces of Python software: BlackBOX \citep{blackbox} and ZOGY\footnote{\href{https://github.com/pmvreeswijk/ZOGY/blob/main/zogy.py}{https://github.com/pmvreeswijk/ZOGY/blob/main/zogy.py}} \citep{Zackay_2016,zogy}. Routine CCD reduction procedures are carried out by BlackBOX on the raw science images, and then ZOGY performs source identification, photometry, astrometry, and finds transient candidates. The method for finding transient candidates is based on the image subtraction routine developed by \citet{Zackay_2016}. The reduced images are available on \texttt{ilifu}, which is a cloud facility\footnote{\href{https://www.ilifu.ac.za/}{https://www.ilifu.ac.za/}} operated by the Inter-University Institute for Data Intensive Astronomy (IDIA) and hosted at the University of Cape Town.

One of MeerLICHT's primary science goals is to try and simultaneously observe FRBs, along with MeerKAT/MeerTRAP. In the early morning on 2023 August 8, MeerLICHT synced with MeerKAT while the latter was observing the MHONGOOSE field J1939--6342, which corresponds to the MeerLICHT fixed sky-grid field 16074. MeerLICHT commenced observations of the field at around 02:05 UT. The arrival time of FRB\,20230808F occurred while the two telescopes were linked, although it was within MeerLICHT's overhead time, in between a \textit{u}-band exposure and a \textit{q}-band exposure. The \textit{q}-band exposure started approximately 3.4 seconds after the arrival time of the FRB, at 03:49:18.32 UT.

The MeerLICHT exposures taken of field 16074 were accessed using \texttt{ilifu}, where we first visually inspected the images taken immediately before and after the arrival of the FRB. No obvious optical emission at the location of the FRB was seen. We created co-adds around the time of the FRB as well, using the co-adding package for MeerLICHT, \texttt{buildref.py}\footnote{\href{https://github.com/pmvreeswijk/BlackBOX/blob/main/buildref.py}{https://github.com/pmvreeswijk/BlackBOX/blob/main/buildref.py}}. The following co-adds were made: a co-add per filter of all MeerLICHT images of field 16074 until 5 days before the FRB (hereafter referred to as the deep co-add), a co-add per filter of all exposures of the field in the 20 minutes before the FRB, one per filter of all exposures in the 20 minutes after the FRB, and then 15 minutes before and after, 5 minutes before and after, and 1 minute before and after the burst arrival time. As the MeerLICHT exposures are 60 seconds in length, the "1-minute" co-adds each had one image only. We then ran ZOGY to perform image subtraction, using the deep co-add for each filter as a reference image, and subtracting these from all of the other co-adds. After performing the image subtraction, we ran the forced photometry package for MeerLICHT, \texttt{force\_phot.py}\footnote{\href{https://github.com/pmvreeswijk/ZOGY/blob/main/force_phot.py}{https://github.com/pmvreeswijk/ZOGY/blob/main/force\_phot.py}} (Vreeswijk et al., in prep), on the resulting images in order to extract flux measurements at the coordinates of the FRB at various times around the arrival of the burst. Photometric and astrometric calibrations for MeerLICHT data are carried out using a catalogue of Gaia DR2 \citep{Gaia_2018} stars.

MeerLICHT re-observed field 16074 on the night of 2023 September 6. We chose a 300\,s integration time, as opposed to the usual 60 seconds, so that we could obtain deeper exposures of the field. These images -- five each in the $u$-, $q$-, $i$- and $z$- bands and four in the $g$-band -- were then also co-added and we ran image subtraction and forced photometry on the resulting co-added images. In addition, field 16074 was given high priority status in MeerLICHT's scheduler, so that it would be observed more frequently as part of MeerLICHT's backup program whenever it was not synced with MeerKAT. As a result of this, the field was observed regularly over the subsequent months, and we created a co-add per filter of all of the MeerLICHT exposures of the field in the 3 months post-FRB, as well as "per night" co-adds for all of the individual nights post-FRB on which the field was observed.

We corrected for Milky Way (MW) dust extinction for all of our photometry results. We found the colour excess in the MW in the direction of FRB\,20230808F, $E(B-V)_{\mathrm{MW}}$, in each of MeerLICHT's filters, based on the color excess map by \cite{Schlafly}\footnote{\href{https://www.ipac.caltech.edu/doi/ned/10.26132/NED5}{https://www.ipac.caltech.edu/doi/ned/10.26132/NED5}}. We assumed the \cite{Fitzpatrick} reddening law with $R_{V}=3.1$, where $R_{V}$ is the total-to-selective extinction ratio. The MW-dust-extinction-corrected photometry results can be seen in Table \ref{tab:Results_Extinct}, and are also shown in Figures~\ref{fig:sub1phot} and \ref{fig:sub2phot}.

\begin{figure}
	\includegraphics[width=\columnwidth]{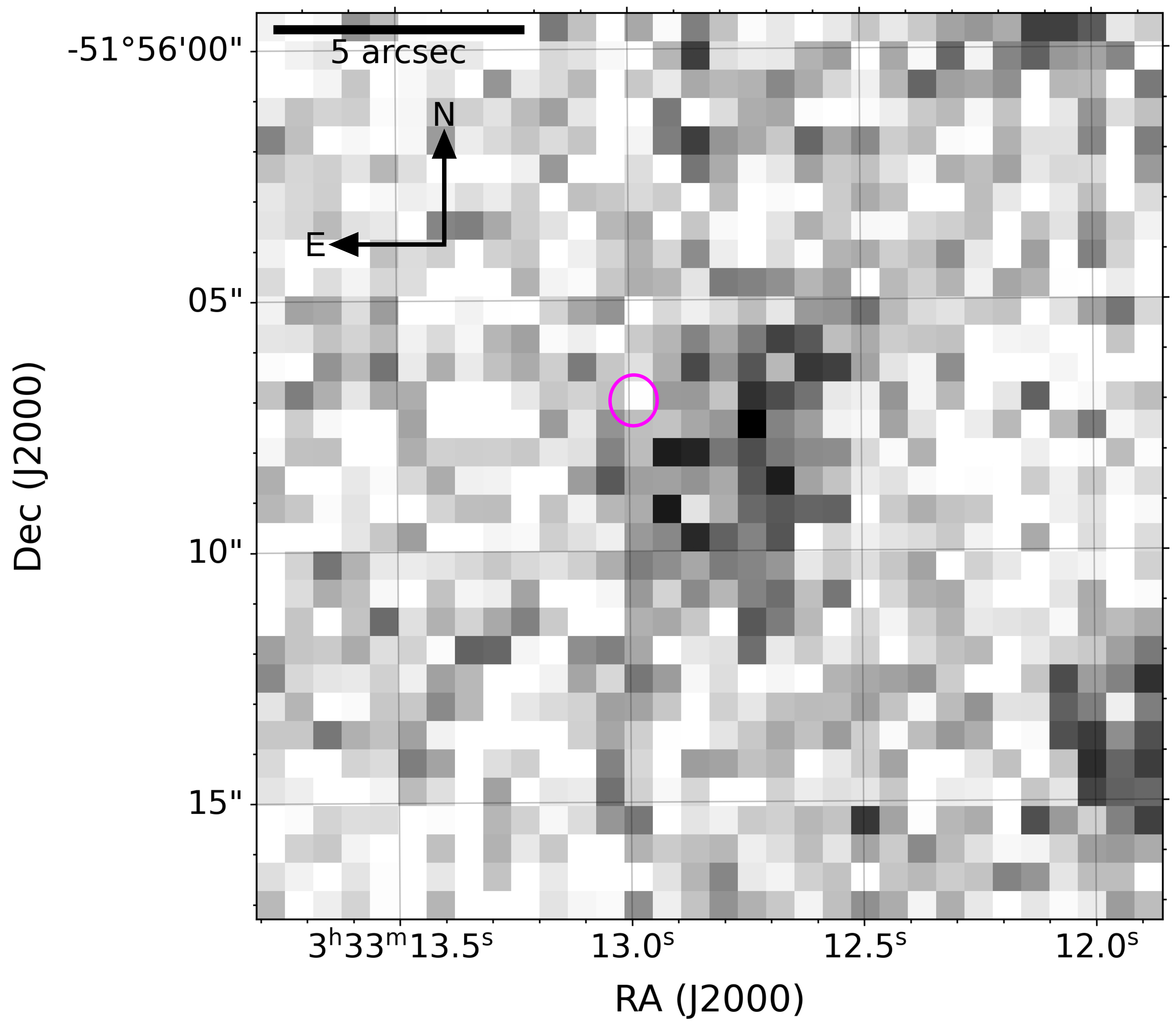}
    \caption{Co-add image produced using all MeerLICHT data of the field of FRB\,20230808F up until shortly before the burst. The magenta ellipse indicates the position of the FRB, and the underlying galaxies are visible but unresolved in this image. The resolved galaxies are shown in Figure~\ref{fig:DES}.}
    \label{fig:ML}
\end{figure}

Our deep co-add, shown in Figure~\ref{fig:ML}, shows a faint underlying extended source at the coordinates of the FRB. This is consistent with the north and south galaxies identified in Section \ref{sec:host_gal}.

\subsection{SALT spectra}\label{sec:SALT}
We observed the two galaxies using the Robert Stobie Spectrograph (RSS; \citealt{rss}) on the 11\,m Southern African Large Telescope (SALT; \citealt{SALT_2006}), to determine their spectroscopic redshifts.

We observed the field of FRB\,20230808F on four separate occasions, and the summary of the observations is shown in Table \ref{tab:SALT-obs}.

\begin{table}
\caption{Observation parameters for the four SALT observations.}
\begin{center}
\begin{tabular}{p{2.5cm} >{\centering\arraybackslash}p{2cm}>{\centering\arraybackslash}p{3.2cm}  }
 \toprule
 \bottomrule
 Date & No. exposures & Length of each exposure (s)\\
 \toprule
 % \bottomrule
 2023 November 3 &2 &1200\\
 2023 November 5 &2 &1200\\
 2023 December 11 &2 &1800 \\
 2023 December 16 &2 &1700\\
 \hline
\end{tabular}
\label{tab:SALT-obs}
\end{center}
\end{table}

For our two observations in November, the SALT Imaging Camera (SALTICAM), SALT's sensitive acquisition camera, had been removed for repairs, and a backup CCD camera, BCAM, was in its place. BCAM is a basic Apogee CCD camera, and is much less sensitive than SALTICAM. The position angle (PA) of the slit was set to $-40$\degree, and we used a slit width of $1.5''$. We had initially requested a slit position angle of $-25$\degree, to pass through both galaxy nuclei as well as two faint alignment stars, but the alignment stars were too faint for BCAM, and as such, we had to adjust the PA to allow for a much brighter alignment star. SALT took these two observations using the PG0700 grating, which covers a wavelength range of 320–900\,nm. The obtained spectra were faint in the continuum, but with a clear emission line at \SI{5021}{\angstrom}.

For the observations on 2023 December 11, the same spectral setup as before was used, but this time with SALTICAM, and therefore the originally desired slit PA of $-25\degree$ was used. The final observations on the 2023 December 16 again used SALTICAM, but this time with a redder spectral set-up, using a slit width of $1.25''$ and the PG0900 grating, which also spans 320--900\,nm but has higher usable angles, increasing the spectral resolution and efficiency at longer wavelengths. A possible emission line was observed at \SI{8840}{\angstrom}, although the spectrum was heavily impacted by skyline residuals. The slit orientations for all of the SALT observations are shown in Figure~\ref{fig:DES}.

\begin{figure}
	\includegraphics[width=\columnwidth]{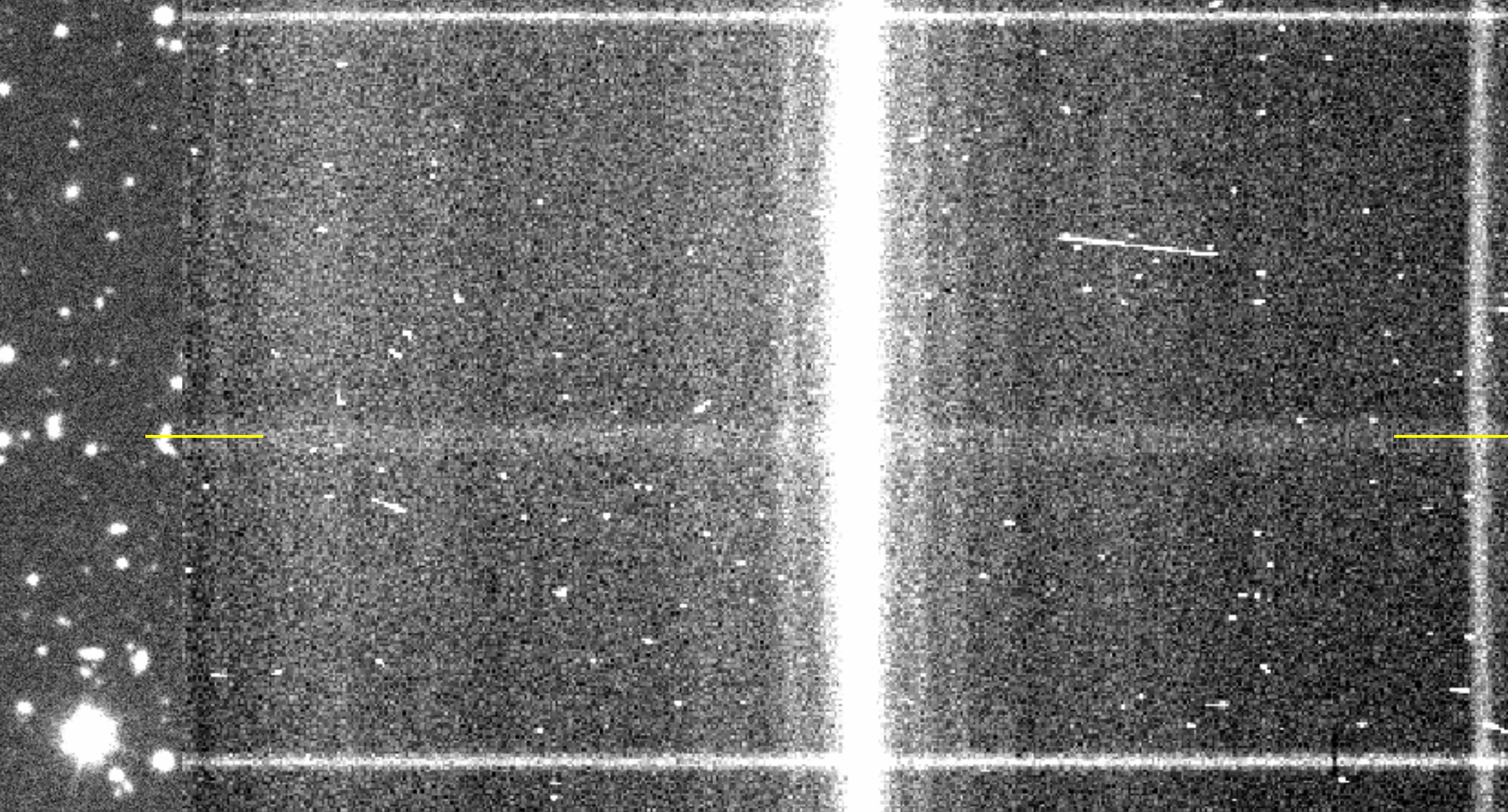}
    \caption{2D trace from a SALT observation on 2023 December 11, with the DES DR2 image of the host galaxies positioned alongside to the left. The two reference stars used on that night are aligned with their respective traces, and two yellow line segments are aligned with the centre of the faint trace, from which we obtained the spectrum. The left line segment passes through the north galaxy, indicating that the spectrum obtained is associated with the north galaxy.}
    \label{fig:Trace}
\end{figure}

\begin{figure}
    \centering
    \begin{subfigure}[t]{0.64\textwidth}
        \centering
        \includegraphics[width=\textwidth]{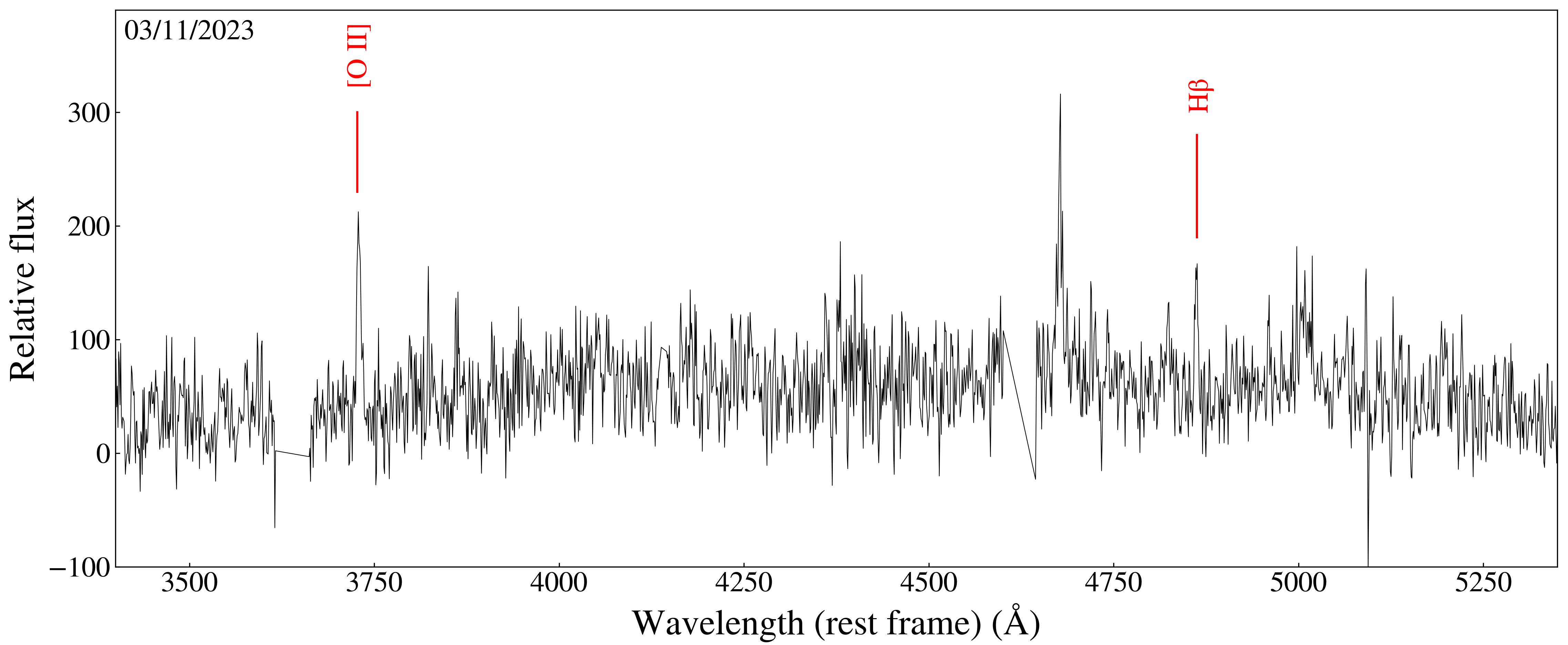}
        \caption{}
        \label{fig:sub1}
    \end{subfigure}
    \hfill
    \begin{subfigure}[t]{0.64\textwidth}
        \centering
        \includegraphics[width=\textwidth]{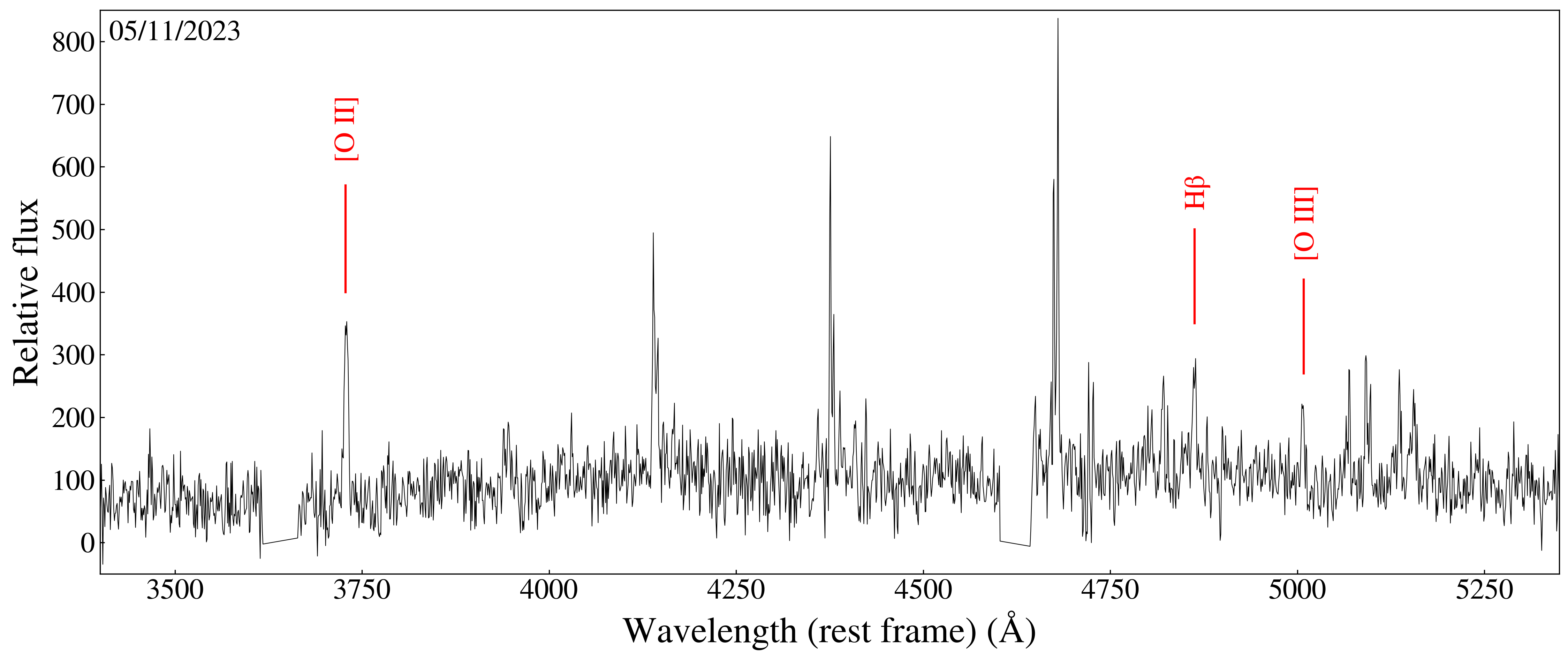}
        \caption{}
        \label{fig:sub2}
    \end{subfigure}
    % \vskip\baselineskip
    \begin{subfigure}[t]{0.64\textwidth}
        \centering
        \includegraphics[width=\textwidth]{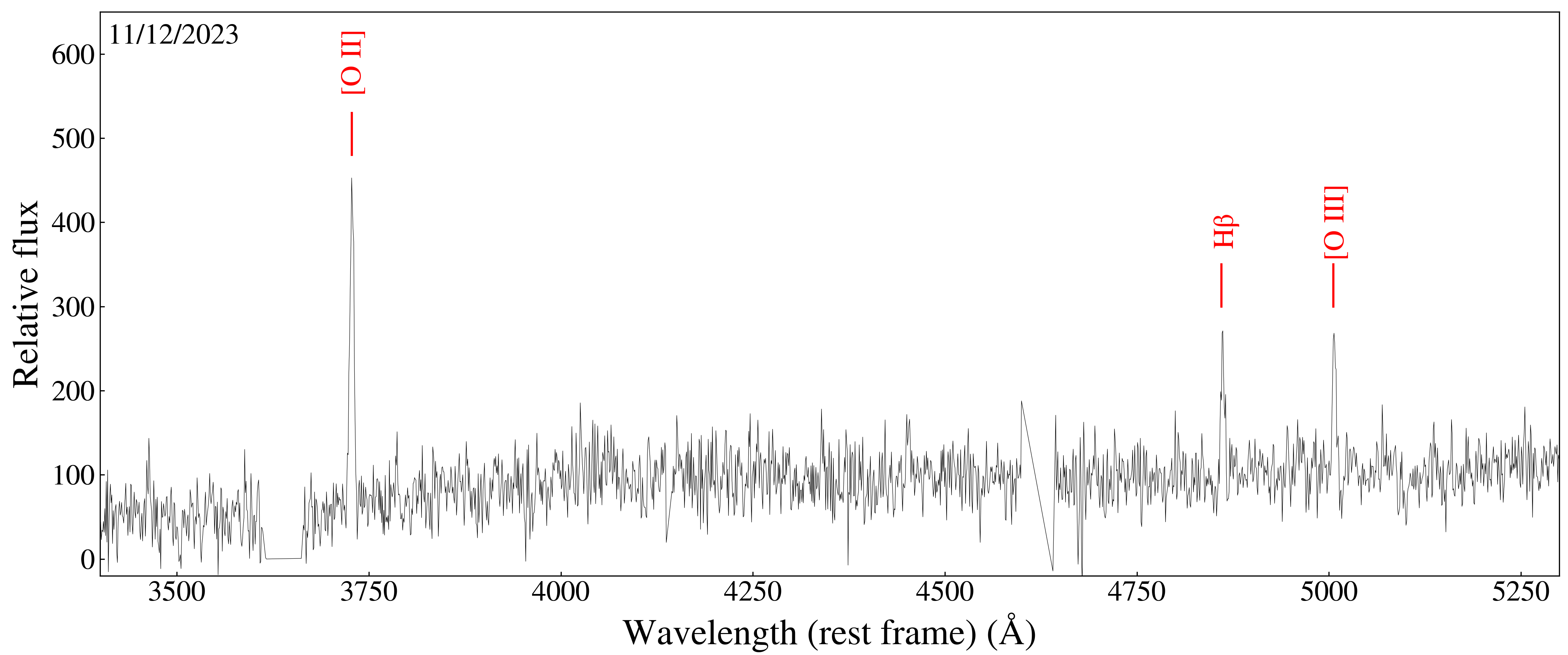}
        \caption{}
        \label{fig:sub3}
    \end{subfigure}
    \hfill
    \begin{subfigure}[t]{0.64\textwidth}
        \centering
        \includegraphics[width=\textwidth]{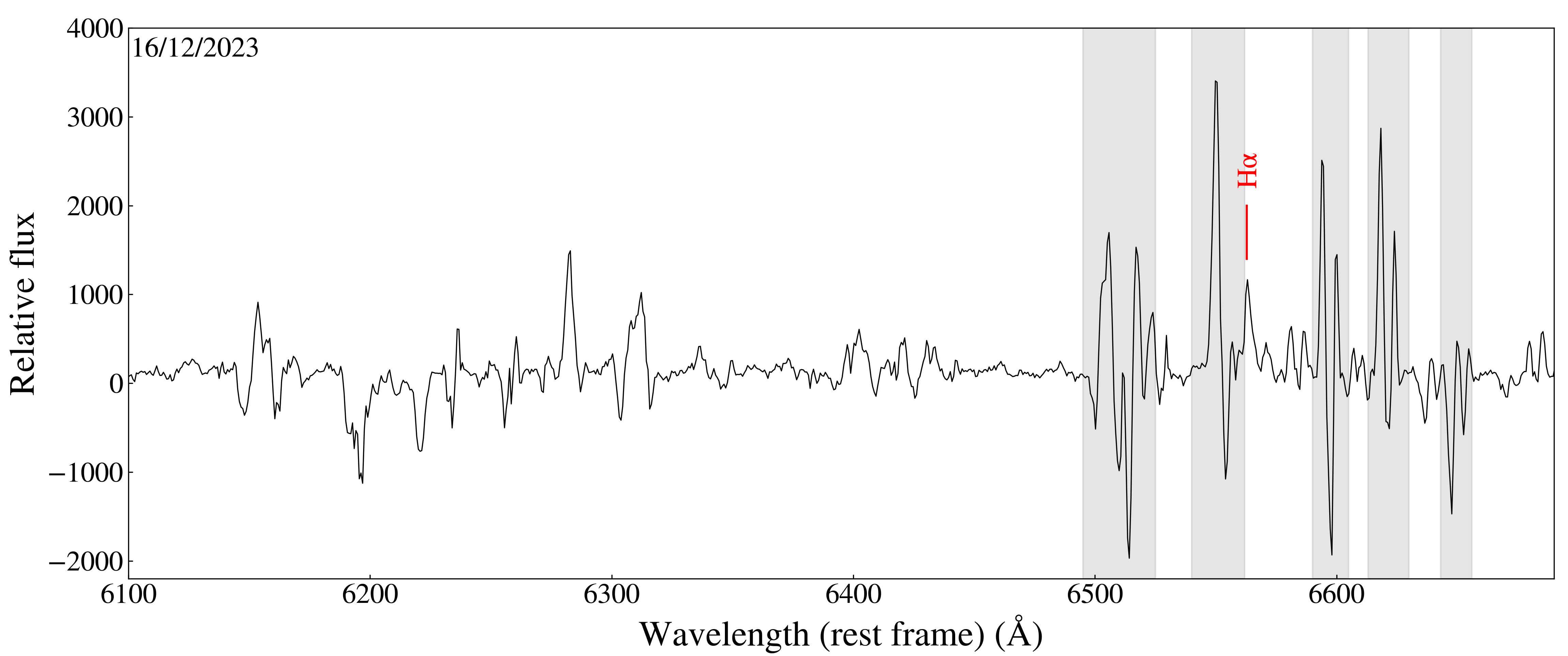}
        \caption{}
        \label{fig:sub4}
    \end{subfigure}
    \caption{SALT spectra obtained over the course of four nights in late 2023. (a) Shows the average combined spectrum from the two observations on 2023 November 3. The unresolved [O II] doublet and H$\beta$ lines are shown by the red lines at \SI{3727}{\angstrom} and \SI{4862}{\angstrom}, respectively. (b) Shows the average combined spectrum from the two observations on 2023 November 5. The [O II], H$\beta$ and [O III] lines are all shown in red, at \SI{3727}{\angstrom}, \SI{4862}{\angstrom} and \SI{5008}{\angstrom}, respectively. (c) Shows the average combined spectrum from the two observations on 2023 December 11. The [O II], H$\beta$ and [O III] lines are all shown in red, at \SI{3727}{\angstrom}, \SI{4862}{\angstrom} and \SI{5008}{\angstrom}, respectively. (d) Shows the average combined spectrum from the two observations on 2023 December 16. The H$\alpha$ emission line is indicated by a red line and the wavelength ranges that have been affected by sky lines are highlighted in grey.}
    \label{fig:main}
\end{figure}

The emission line at \SI{5021}{\angstrom} was determined to be the [O II] doublet, which corresponds to a redshift of $z_{\mathrm{spec}}\sim$0.347, and there are confirming H$\beta$ and [O III] lines at the same redshift at \SI{6548}{\angstrom} and \SI{6743}{\angstrom}, respectively. All three lines are clearly visible in the exposures from 2023 November 5 and 2023 December 11, while the spectrum from 2023 November 3 shows the [O II] and H$\beta$ emission lines. The emission line at \SI{8840}{\angstrom} from 2023 December 16 matches the expected position of H$\alpha$ at the calculated redshift. The spectra from all four nights are shown, with their identified emission lines labeled in the rest frame, in Figures~\ref{fig:sub1}, \ref{fig:sub2}, \ref{fig:sub3} and \ref{fig:sub4}. The emission in all of the spectra was determined to be from the north galaxy by comparing the distances between the traces in the 2D spectra and the distances between the reference stars and the galaxies in the various observation set-ups. An example of this is shown in Figure~\ref{fig:Trace}.

The calculated spectroscopic redshift of $z_{\mathrm{spec}}=0.3472\pm0.0002$ is in agreement with the photometric redshift of the north galaxy within $1.13\,\mathrm{\sigma}$. Using the spectroscopic redshift, we calculated the luminosity distance, $D_L$, as $1903.4\,\mathrm{Mpc}$, assuming cosmological parameters from the \cite{Planck_2020}.

\section{FRB 20230808F characteristics}
\subsection{Radio flux and fluence}\label{sec:Radio flux}
The peak flux density for FRB\,20230808F was calculated as described in \cite{jankowski_flux}, using a modified version of the single-pulse radiometer equation \citep{Dewey_1985}:
\begin{equation}
    S_{\mathrm{peak}}(\mathrm{S/N,}W_{\mathrm{eq}},\vec{a})=\mathrm{S/N \:\beta \:\eta_{b}}\:\frac{T_{\mathrm{sys}}+T_{\mathrm{sky}}}{G\sqrt{b_{\mathrm{eff}}N_{\mathrm{p}}W_{\mathrm{eq}}}}\:a^{-1}_{\mathrm{CB}}\:a^{-1}_{\mathrm{IB}},
\end{equation}
where $S_{\mathrm{peak}}$ is the peak flux density, $\vec{a}$ is the parameter vector, $\beta$ is the digitisation loss factor, which is 1.0 for 8-bit digitisation \citep{Kouwenhoven_2001}, $\eta_{\mathrm{b}}$ is the beam-forming efficiency, taken to be $\sim$1 \citep{Chen_2021}, $G$ is the telescope forward gain, $b_{\mathrm{eff}}$ is the effective bandwidth, $N_{\mathrm{p}}=2$ is the number of polarisations summed, $W_{\mathrm{eq}}$ is the observed equivalent boxcar pulse width, $T_{\mathrm{sys}}=19\,\mathrm{K}$ and $T_{\mathrm{sky}}$ are the system and sky temperatures, and $a_{\mathrm{CB}}=0.9$ and $a_{\mathrm{IB}}=0.8$ are the attenuation factors of the detection CB and the IB. The radio flux density was thus found to be $S_{\mathrm{peak}}=(157\pm14)$\,mJy, calculated from the filterbank data where the FRB was detected. The fluence $F_{\mathrm{radio}}$ was then calculated from $F_{\mathrm{radio}}=S_{\mathrm{peak}}W_{\mathrm{eq}}$, yielding $F_{\mathrm{radio}}=(1.72\pm0.01)\,\mathrm{Jy\,ms}$.

Using the peak flux density and fluence, we computed the isotropic luminosity, $L_{p}$, and energy, $E$, of the burst, as described in \cite{ZhangBing_2018}, using:
\newpage
\begin{equation}
    L_{p}\simeq4\pi D_{L}^2S_{\nu,p}\nu_{c}=(10^{42}\,\mathrm{erg\,s^{-1}})\,4\pi\left(\frac{D_{\mathrm{L}}}{10^{28}\,\mathrm{cm}}\right)^{2}\frac{S_{\nu,p}}{\mathrm{Jy}}\frac{\nu_{c}}{\mathrm{GHz}},
\end{equation}

\begin{equation}
    E\simeq\frac{4\pi D_{\mathrm{L}}^{2}}{1+z}F_{\nu}\nu_{c}=(10^{39}\,\mathrm{erg})\frac{4\pi}{1+z}\left(\frac{D_L}{10^{28}\mathrm{cm}}\right)^{2}\frac{F_{\nu}}{\mathrm{Jy\cdot ms}}\frac{\nu_c}{\mathrm{GHz}},
\end{equation}
where $S_{\nu,p}$ is the specific peak flux density, $F_{\nu}$ is the specific fluence, $\nu_c$ is the central frequency of the observing band, and $D_L$ is the luminosity distance, obtained in Section~\ref{sec:SALT}. Using the values pertaining to FRB\,20230808F, we obtained $L_{p}=8.7\times10^{41}\,\mathrm{erg\,s^{-1}}$ and $E=7.1\times10^{39}\,\mathrm{erg}$. However, we note that using the central frequency when calculating the luminosity and energy would imply a flat burst spectrum, and as such, following the prescription of \cite{Aggarwal_2021}, we also used the burst bandwidth, instead of the central observing frequency, to calculate the energy and luminosity of FRB\,20230808. Doing so yielded $L_{p}=2.0\times10^{41}\,\mathrm{erg\,s^{-1}}$ and $E=1.6\times10^{39}\,\mathrm{erg}$.

\subsection{Host galaxy contribution to propagation properties}
We used the \texttt{FRBs}\footnote{\href{https://github.com/FRBs/FRB}{https://github.com/FRBs/FRB}} library \citep{Prochaska_2019, j_xavier_prochaska_2023_8125230} to calculate the expected DM contributions, rotation measure, and redshift for FRB 20230808.

The total observed FRB DM is the sum of the DM contributions from the MW, MW halo, the intergalactic medium (IGM), and the host galaxy, and can thus be expressed as 
\begin{equation}
    \mathrm{DM=DM_{MW}+DM_{halo}+DM_{IGM}+\frac{DM_{host}}{1+z}}.
\end{equation}
The expected MW DM contribution in the direction of the FRB, according to the YMW16 model \citep{Yao_2017}, is $\sim$$35\,\mathrm{pc\,cm^{-3}}$, with an expected  $\tau_{s}\sim4.8\times10^{-8}$\,s in the $L$-band. According to the NE2001 model \citep{Cordes_2003}, the MW DM contribution is $\sim$$27\,\mathrm{pc\,cm^{-3}}$, with an expected  $\tau_{s}\sim4.3\times10^{-8}$\,s in the $L$-band. Taking the average between the YMW16 and NE2001 $\mathrm{DM_{MW}}$ values, we obtain $\mathrm{DM_{MW}\sim31\,pc\,cm^{-3}}$. We used the model from \cite{Yamasaki_2020} to estimate the DM contribution from the MW halo ($\mathrm{DM_{halo}}$), which was found to be $\sim$$33\,\mathrm{pc\,cm^{-3}}$.

The redshift upper limit was computed using the cosmological parameters from the \cite{Planck_2020}. By applying the Macquart relation \citep{Macquart_2020}, we adjusted the value of $\mathrm{DM_{host}}$ until the predicted redshift matched the spectroscopic redshift of $0.3472\pm0.0002$, thus obtaining $\mathrm{DM_{host}\sim420\,pc\,cm^{-3}}$ in the host galaxy reference frame. Subtracting all of the calculated DM contributions from the total observed DM of $\sim$$653\,\mathrm{pc\,cm^{-3}}$, we obtain $\mathrm{DM_{IGM}\sim277\,pc\,cm^{-3}}$.

We calculated the MW contribution to the RM using the Faraday sky model from \cite{Hutschenreuter_2020}. The expected MW contribution to the RM in the direction of the FRB is $\sim$$18\,\mathrm{rad\,m^{-2}}$, while that at the host was found to be $\sim$$274\,\mathrm{rad\,m^{-2}}$.

The expected MW scattering timescales are orders of magnitude smaller than the actual FRB scattering timescale of $\tau_s = (3.10\pm0.13)\,\mathrm{ms}$. The expected scattering timescales and the calculated values for the DM and RM contributions, indicate that there are significant extragalactic contributions to all of these propagation properties from either the FRB environment, the host galaxy, or a foreground galaxy. However, disentangling the individual contributions from each of these components is challenging.

\subsection{Optical flux limits}
\begin{figure}
    \centering
    \begin{subfigure}[b]{1.0\textwidth}
        \centering
        \includegraphics[width=\textwidth]{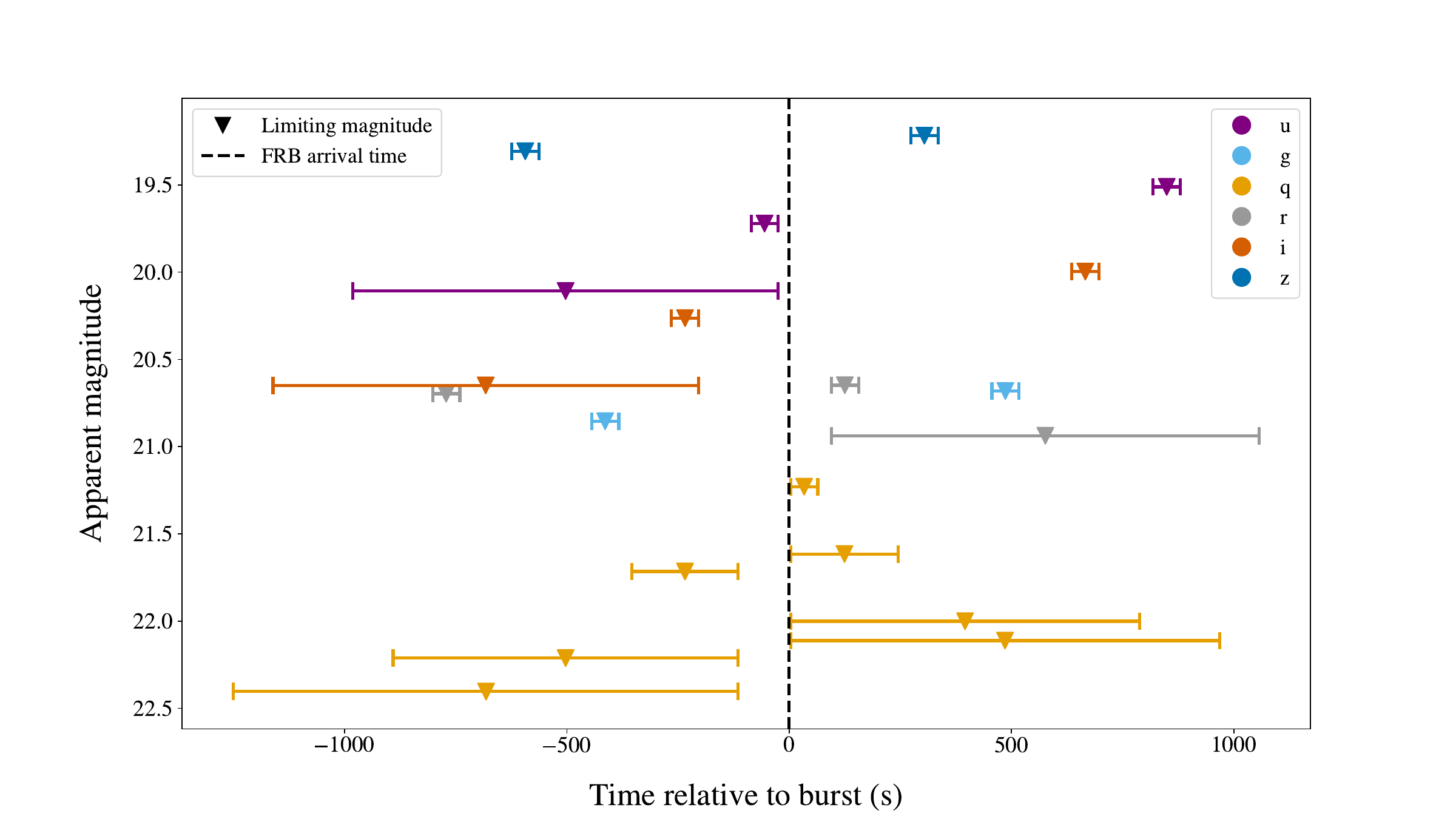}
        \caption{}
        \label{fig:sub1phot}
    \end{subfigure}
    \vskip\baselineskip
    \begin{subfigure}[b]{1.0\textwidth}
        \centering
        \includegraphics[width=\textwidth]{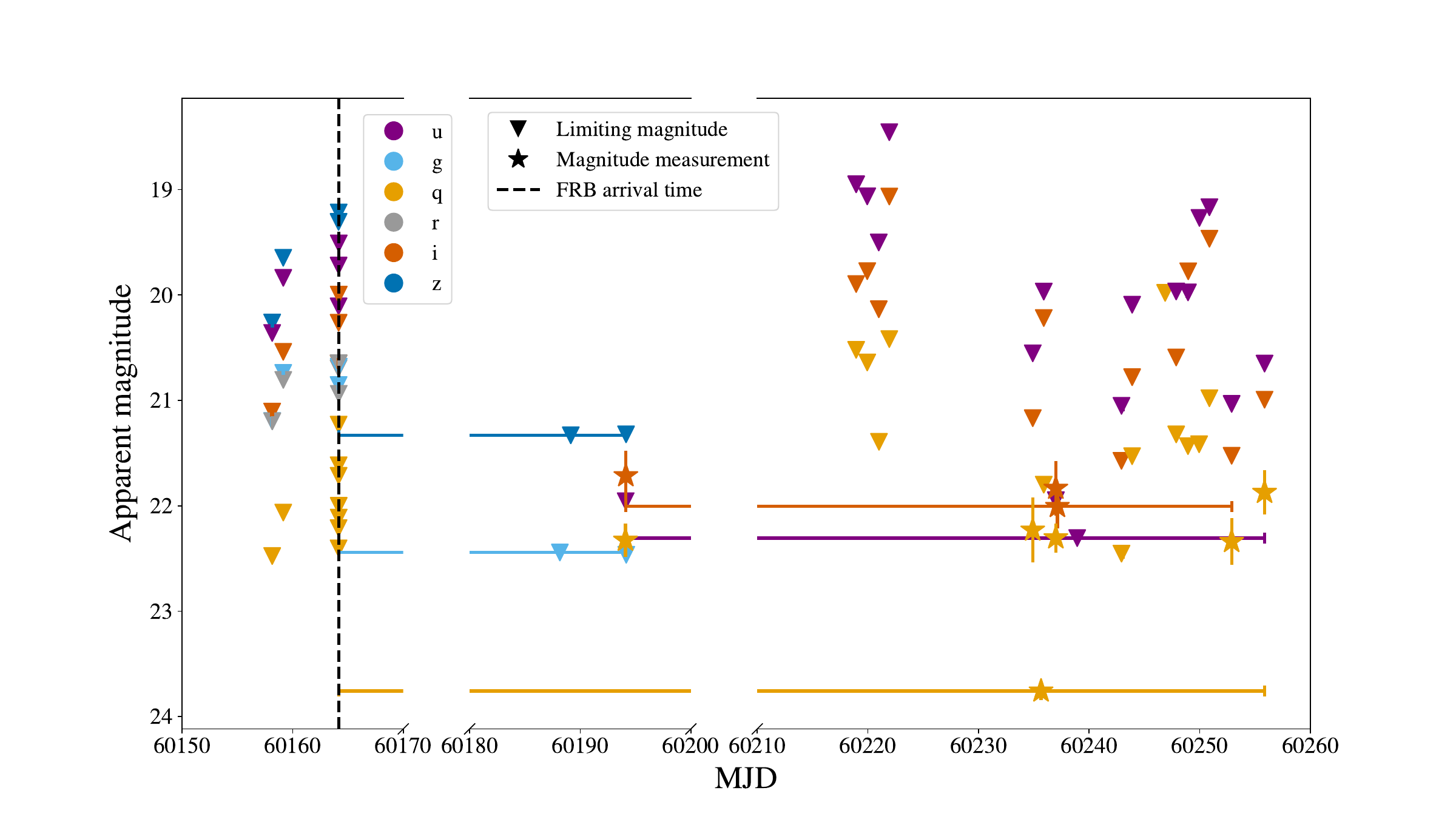}
        \caption{}
        \label{fig:sub2phot}
    \end{subfigure}
    \caption{Forced photometry results for FRB\,20230808F. In both plots, each colour represents a different MeerLICHT filter, upside down triangles indicate limiting magnitudes, and stars represent actual magnitude measurements where the S/N was greater than 3. The dashed black line in each plot indicates the MJD at which FRB\,20230808F took place. The forced photometry results from the night of the FRB, with apparent magnitude plotted against modified Julian date (MJD), and horizontal errorbars representing the time-span over which a co-add was made, are shown in (a). The forced photometry results at the FRB coordinates, for all of the MeerLICHT images of field 16074, but excluding the deep co-add, are shown in (b). The deep co-add is excluded because it was used as the reference image for image subtraction. For upper limits obtained from co-added images spanning a short time range, the horizontal error bars are not visible in (b).}
    \label{fig:mainphot}
\end{figure}

Table \ref{tab:Results_Extinct} summarizes the upper limits for optical emission in the seconds, minutes and months following the FRB arrival time. For many of the co-adds we created, the S/N was lower than 3, and for these cases we used the limiting magnitude at a significance level of $3\sigma$ at the coordinates of the FRB instead of the magnitude value and associated error. As such, in Table \ref{tab:Results_Extinct}, the limits which include  uncertainties are those where the S/N was greater than 3, and hence where the actual magnitude value was used. In such cases, we assume that the measurement is of the underlying host galaxy. Figures~\ref{fig:sub1phot} and \ref{fig:sub2phot} show plots of the upper limits derived from the photometry. Points on these plots which have vertical errorbars are those for which actual magnitude measurements were used, whereas points without represent limiting magnitudes. The horizontal "error bars" represent the timespans over which the images used to create the co-adds were obtained.

No optical counterpart was found for FRB\,20230808F in our almost simultaneous or follow-up observations. In the time just after the burst, we were able to set an optical upper flux limit in the $q$-band of $f_{\mathrm{AB,}q}<11.7\,\micro \mathrm{Jy}$, while for our $u$-band observation which ended approximately 24.7 seconds before the FRB arrival time, we obtained an upper limit of $f_{\mathrm{AB,}u}<46.9\,\micro \mathrm{Jy}$ We also obtained upper limits on various nights over the course of the 3 months after the FRB.

\onecolumn
\twocolumn

\begin{table}
\caption{DESI DR10 magnitudes for the two potential identified host galaxies.}
\begin{center}
\begin{tabular}{ >{\centering\arraybackslash}p{1cm}>{\centering\arraybackslash}p{2cm}>{\centering\arraybackslash}p{2cm}  }
 \toprule
 \bottomrule
 Filter & North galaxy & South Galaxy\\
 \toprule
 % \bottomrule
 $g$ &22.3 &22.9\\
 $r$ &21.4 &21.6\\
 $i$ &21.0 &21.1 \\
 $z$ &20.6 &20.7\\
 \hline
\end{tabular}
\label{tab:DESI_mags}
\end{center}
\end{table}

\begin{table}
\caption{Estimated age, log stellar mass, and star formation rate for each of the potential host galaxies, obtained from the SED fitting performed by \protect\cite{Zou_2022}. The upper SFR refers to the upper limit of the star formation rate with a $68$ per cent confidence level.}
\begin{center}
\begin{tabular}{p{2.5cm} >{\centering\arraybackslash}p{2cm}>{\centering\arraybackslash}p{2cm}  }
 \toprule
 \bottomrule
  Property & North galaxy & South galaxy\\
 \toprule
 % \bottomrule
 Age (yr)  & $8\times10^9$ & $4\times10^9$ \\
 Mass (dex) & 10.26 & 10.33 \\
 Best SFR ($M_{\odot}\,\mathrm{yr^{-1}}$) & $-0.1$ & $-0.1$ \\
 Upper SFR ($M_{\odot}\,\mathrm{yr^{-1}}$) & $0.2$ & $0.0$ \\
 \hline
\end{tabular}
\label{tab:SED_prop}
\end{center}
\end{table}

\onecolumn
\twocolumn

\subsection{Fluence ratio limit and optical luminosity limit}\label{sec:Luminosity}
The flux upper limit in the $q$-band of $f_{\mathrm{AB,}q}<11.7\,\micro \mathrm{Jy}$ corresponds to a fluence of $F_{\mathrm{opt}}=0.039\,\mathrm{Jy\,ms}$ if we conservatively assume that the duration of a possible optical burst $t_{\mathrm{opt}}\approx\Delta t$, where $\Delta t\approx3.4\,\mathrm{s}$ is the time delay between the FRB and the start of our optical observation. Choosing this timescale allows us to consider the shortest possible optical burst that could still be detectable under our observing conditions, and this is the same timescale used by \cite{Hiramatsu_2023} for calculating fluence limits where there was a delay between the burst arrival time and the start of an optical exposure. We thus obtain a fluence ratio $F_{\mathrm{opt}}/F_{\mathrm{radio}}\lesssim0.023$ on a timescale of $\sim$$3.4\,\mathrm{s}$. We also evaluated the fluence limit on a longer timescale of the delay time plus the full integration time. In this case, the fluence limit is $F_{\mathrm{opt}}=0.74\,\mathrm{Jy\,ms}$, and the fluence ratio is then $F_{\mathrm{opt}}/F_{\mathrm{radio}}\lesssim0.43$ on a timescale of $\sim$$63.4\,\mathrm{s}$.

For our first observation after the FRB, we were able to obtain a MeerLICHT $q$-band luminosity limit of $\nu L_{\nu}\sim1.3\times10^{43}\,\mathrm{erg\,s^{-1}}$ with a 60\,s exposure starting $\sim$3.4\,s after the arrival of the FRB. This corresponds to an optical energy limit $E_{\mathrm{opt}}\sim7.9\times 10^{44}$\,erg, which yields an energy ratio $E_{\mathrm{opt}}/E_{\mathrm{radio}}\lesssim 4.8\times10^{5}$ if we use the radio burst energy calculated using the burst bandwidth.

\section{FRB 20230808F in context}
\subsection{Comparison to previous optical limits}
\cite{Hardy_2017} simultaneously observed the repeater FRB\,20121102, which has a redshift of $z=0.19$ \citep{Tendulkar_2017}, with the 100-m Effelsberg Radio Telescope and ULTRASPEC, which is mounted on the 2.4-m Thai National Telescope \citep{Ultraspec}. They detected 13 bursts in the radio and stacked the optical data around the arrival time of each of the bursts. They obtained an optical fluence limit of $0.17\,\mathrm{Jy\,ms}$ on a timescale of  $141.4$\,ms, corrected for MW dust extinction, which corresponds to an optical-to-radio fluence ratio limit of $F_{\mathrm{opt}}/F_{\mathrm{radio}}\lesssim0.28$.

The \cite{Magic_2018} conducted simultaneous radio, optical, and very-high-energy (VHE) gamma-ray observations of FRB\,20121102, making use of the Arecibo radio telescope and the Major Atmospheric Gamma Imaging
Cherenkov (MAGIC) telescope, located on La Palma in the Canary Islands. The MAGIC telescopes are primarily designed for VHE gamma-ray detection, but can also perform optical observations using the central pixel, allowing for simultaneous optical and gamma-ray measurements during their operations. The authors detected five radio bursts and stacked the optical data around the arrival times of the bursts, obtaining optical fluence limits of $0.025\,\mathrm{Jy\,ms}$ on a timescale of $0.1\,\mathrm{ms}$, $0.086\,\mathrm{Jy\,ms}$ on a timescale of $1\,\mathrm{ms}$, $0.25\,\mathrm{Jy\,ms}$ on a timescale of $5\,\mathrm{ms}$, and $0.36\,\mathrm{Jy\,ms}$ on a timescale of $10\,\mathrm{ms}$. All of these limits were corrected for MW dust extinction and correspond to optical-to-radio fluence ratio limits of $F_{\mathrm{opt}}/F_{\mathrm{radio}}\lesssim0.013-0.18$.

\cite{Niino} used Tomo-e Gozen \citep{Tomo-e}, a high-speed camera mounted on the Kiso 105-cm Schmidt telescope, located at Kiso Observatory in Japan, to conduct optical observations of the repeater FRB\,20190520B, which has a redshift of $z = 0.24$ \citep{Niu_2022}, while simultaneously obtaining radio observations with the Five-hundred-meter Aperture Spherical radio
Telescope (FAST). While they detected 11 radio bursts, no corresponding optical emission was found. They obtained MW-dust-extinction-corrected optical fluence limits as deep as $0.068\,\mathrm{Jy\,ms}$ on a timescale of $40.9$\,ms for the individual bursts. A limit of $0.029\,\mathrm{Jy\,ms}$ was obtained by averaging the optical flux of the 11 bursts. They also investigated the optical fluence on longer timescales, obtaining a fluence limit of $0.13\,\mathrm{Jy\,ms}$ on a timescale of $0.45$\,s, and $0.34\,\mathrm{Jy\,ms}$ on a timescale of $2.1$\,s. Their limits correspond to $F_{\mathrm{opt}}/F_{\mathrm{radio}}\lesssim0.23-2.1$ for individual bursts, and $F_{\mathrm{opt}}/F_{\mathrm{radio}}\lesssim0.20$ for the stacked optical limit.

\cite{Hiramatsu_2023} conducted simultaneous, follow-up, and serendipitous archival survey observations of 15 well-localised FRBs, consisting of 7 repeating FRBs, 7 non-repeaters, and SGR 1935+2154. Their almost simultaneous observation with a 15\,s integration time of the repeater FRB\,20220912A, which has a redshift of $z=0.077$ \citep{Ravi_2023} using Binospec \citep{Binospec} on the 6.5 m MMT Observatory in Arizona, USA, set a limit on the fluence ratio of $F_{\mathrm{opt}}/F_{\mathrm{radio}}\lesssim0.00010$ on a timescale of $\sim$ 0.4\,s. They also observed the FRB using KeplerCam \citep{szentgyorgyi2005keplercam} on the 1.2 m Telescope at the Fred Lawrence
Whipple Observatory in Arizona, USA, and the Sinistro camera on the 1 m telescope at the McDonald
Observatory in Texas, USA, in the Las Cumbres Observatory (LCO; \citealt{LCO}) network. They obtained a luminosity limit $\nu L_{\nu}\sim2\times10^{41}\,\mathrm{erg\,s^{-1}}$ for their Binospec observation, while the luminosity limits for the same FRB that they obtain from simultaneous observations using KeplerCam and LCO range from $\sim(0.3-2.9)\times10^{42}\,\mathrm{erg\,\mathrm{s^{-1}}}$, which they report as being the deepest luminosity limits to date for an extragalactic FRB. These simultaneous luminosity limits correspond to fluence ratio limits as deep as $F_{\mathrm{opt}}/F_{\mathrm{radio}}\lesssim0.0010$.

\cite{kilpatrick_2023} used the `Alopeke high-cadence camera \citep{Scott_2018, Scott_2021} on the Gemini North telescope, located on Mauna Kea in Hawaii, to obtain  optical observations of two bursts from the repeater FRB\,20180916B, which is at a redshift of $z=0.034$ \citep{Marcote_2020}. These observations occurred  simultaneously with radio observations conducted by CHIME. They obtained MW-dust-extinction-corrected optical fluence limits on timescales of $\sim$10.4\,ms for each burst, corresponding to optical-to-radio fluence ratio limits of $F_{\mathrm{opt}}/F_{\mathrm{radio}}<0.002-0.007$.

\cite{trudu} conducted a multi-wavelength radio, optical and X/$\gamma$-ray campaign on FRB\,20180916B. They observed the FRB over the course of 8 months, using the Sardinia Radio Telescope (SRT) and the upgraded Giant Metrewave Radio Telescope (uGMRT; \citealt{Gupta_2017}) for their radio observations. For their optical observations, they made use of the Aqueye+ fast photon counter \citep{Barbieri_2009,Zampieri_2015}, mounted on the Copernicus telescope, and the IFI+Iqueye fast photon counter \citep{Naletto_2009}, mounted on the Galileo telescope, both located in Asiago. Additionally, they made use of the Calar Alto 2.2\,m telescope, the 2.5\,m telescope at the Caucasian Mountain Observatory (CMO) of
the Sternberg Astronomical Institute (SAI) Lomonosov Moscow State University (MSU), and the SiFAP2 fast photometer \citep{Ghedina_2018,Ambrosiono_2016} on the 3.6\,m Telescopio Nazionale Galileo (TNG) in the Canary Islands. The deepest simultaneous optical fluence upper limits they obtained with Aqueye+ were $0.005\,\mathrm{Jy\,ms}$ on a timescale of 200\,ms, and $0.009\,\mathrm{Jy\,ms}$ on a timescale of 30\,s in November 2020. In August 2021, they obtained upper limits of $0.005\,\mathrm{Jy\,ms}$ on a timescale of 200\,ms and $0.008\,\mathrm{Jy\,ms}$ on a timescale of 30\,s. Their upper limits from IFI+Iqueye were $0.09\,\mathrm{Jy\,ms}$ on a timescale of 200\,ms and $0.13\,\mathrm{Jy\,ms}$ on a timescale of 30\,s. Their upper limits using TNG were $ 1.02\,\mathrm{mJy\,ms}$ on a timescale of 200\,ms and $1.56\,\mathrm{mJy\,ms}$ on a timescale of 30\,s. None of the upper limits reported by \cite{trudu} were corrected for extinction. They set upper limits on the energy ratio $E_{\mathrm{opt}}/E_{\mathrm{radio}}< 1.3\times10^2-7.8\times10^2$, which they report as being the most stringent upper limits for FRB\,20180916B to date.

\begin{figure}
	\includegraphics[width=\columnwidth]{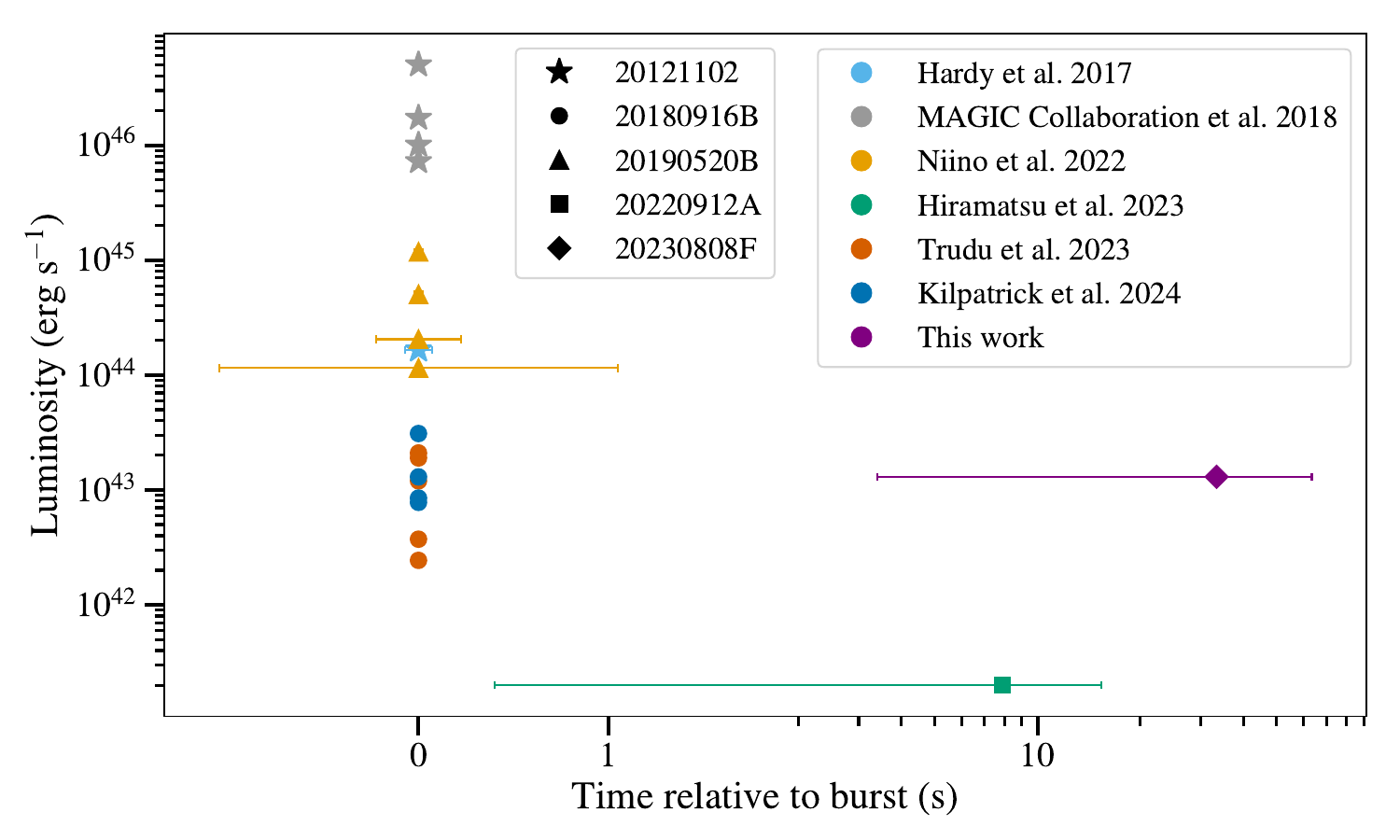}
    \caption{Luminosity limits obtained by \protect\cite{Hardy_2017}, the \protect\cite{Magic_2018}, \protect\cite{Niino}, \protect\cite{Hiramatsu_2023}, \protect\cite{trudu} and \protect\cite{kilpatrick_2023}, along with our luminosity limit. Limits for different FRBs are represented by different shapes, while those from different works are shown by different colours. The x-axis is linear near zero to display upper limits centered on the FRB arrival time, and then transitions to a logarithmic scale to accommodate larger timescales.}
    \label{fig:Luminosities}
\end{figure}

Our fluence limit of $0.039\,\mathrm{Jy\,ms}$ is deeper than the limit obtained by \cite{Hardy_2017} and those obtained by \cite{Niino} for individual bursts and on their timescales of 0.45\,s and 2.1\,s. The fluence ratio obtained by \cite{Hiramatsu_2023} is deeper than ours and is on a shorter timescale. Our fluence limit is deeper than those obtained by the \cite{Magic_2018} on timescales greater than or equal to 1\,ms. The almost simultaneous luminosity limit obtained by \cite{Hiramatsu_2023} is two orders of magnitude deeper than ours. \cite{kilpatrick_2023} also obtained limits deeper than our own and on much shorter timescales. The energy ratio limits obtained by \cite{trudu} are 3 orders of magnitude deeper than our own. Their Aqueye+ and SiFAP2 fluence limits are also deeper than ours on both of their considered timescales, as are their deepest luminosity limits. The fluence ratio limits of the previous works discussed, where available, along with our own limits, are presented in Table \ref{tab:fluence-lims}, and the associated luminosity limits are shown in Figure~\ref{fig:Luminosities}.

We note that the aforementioned optical limits were all for repeater FRBs, whereas ours are for a so far non-repeating burst. \cite{Hiramatsu_2023} did some analysis based on archival optical observations of non-repeating FRBs, but all of the optical observations of one-offs available to them, had delays $\gtrsim$$10^4$\,s with respect to the time of the bursts. \cite{Nunez_2021} were able to obtain optical observations of the non-repeating FRB 20190711 using the Las Cumbres Observatory Global Telescope (LCOGT) network \citep{Brown_2013}, but their 60\,s exposures commenced almost two hours after the FRB's arrival time. Our optical observation time delay of $\sim$3.4\,s is by far the smallest such delay for a non-repeater to date. All of the optical-to-radio fluence ratios for non-repeaters in \cite{Hiramatsu_2023} are $\gtrsim$1, and as such our value of $F_{\mathrm{opt}}/F_{\mathrm{radio}}\lesssim0.023$ is a significant improvement.

\begin{table}
\caption{Fluence ratio limits for previous studies discussed, along with the limit obtained in this work. Aside from our own limit, all of the other limits above are for repeating FRBs.}
\begin{center}
\begin{tabular}{ >{\arraybackslash}p{4cm}>{\arraybackslash}p{2.5cm} >{\centering\arraybackslash}p{0.9cm} }
 \toprule
 \bottomrule
  Work & $F_{\mathrm{opt}}/F_{\mathrm{radio}}$ & Repeater\\
 \toprule
 % \bottomrule
 \cite{Hardy_2017} &$\lesssim0.28$ & Yes\\
 \cite{Magic_2018} &$\lesssim0.013-0.18$ & Yes\\
 \cite{Niino} &$\lesssim0.20-2.1$ & Yes\\
 \cite{Hiramatsu_2023} &$\lesssim0.00010-0.0010$ & Yes\\
 \cite{kilpatrick_2023} &$<0.0020-0.0070$ & Yes\\
 This work &$\lesssim0.023; \lesssim0.43$ & No\\
 \hline
\end{tabular}
\label{tab:fluence-lims}
\end{center}
\end{table}

\subsection{Constraints on the synchrotron maser model}
\begin{figure}
	\includegraphics[width=\columnwidth]{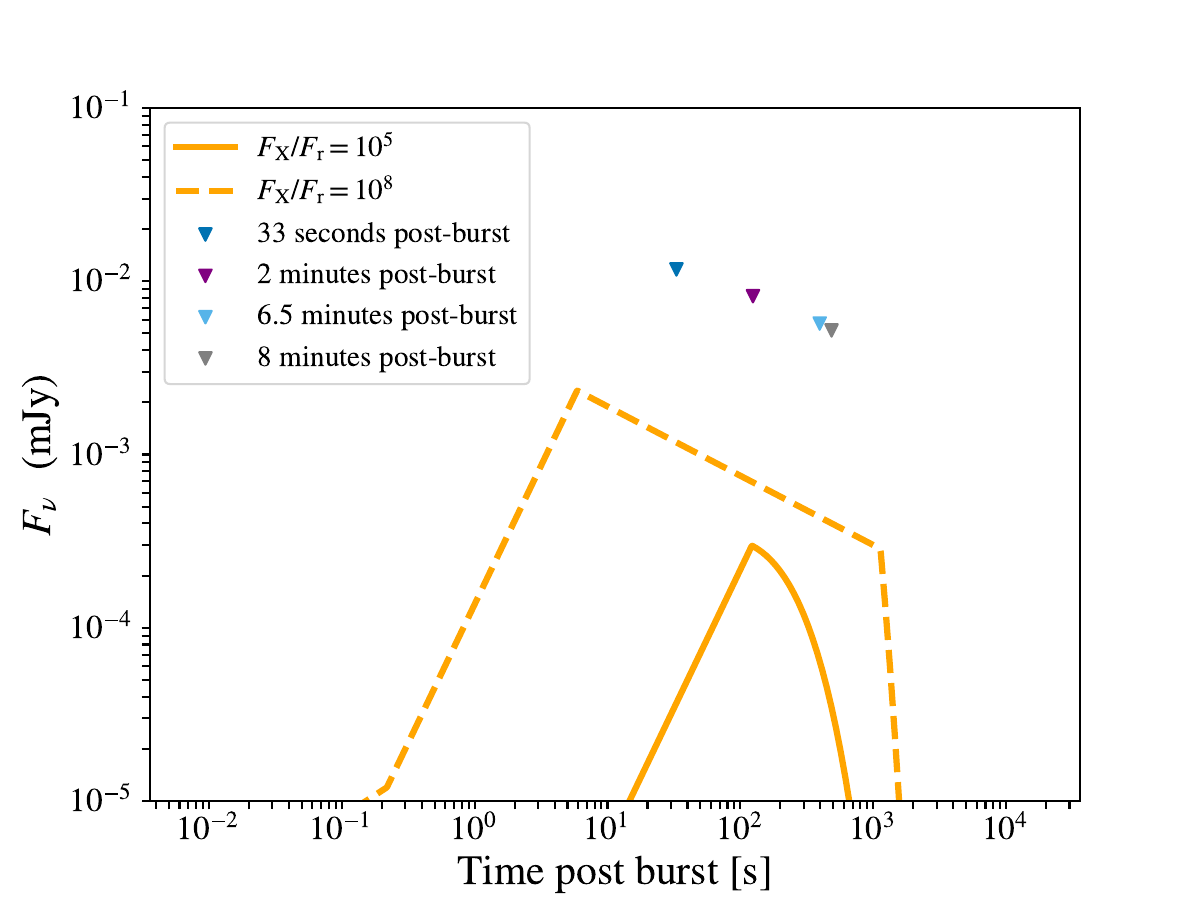}
    \caption{Optical afterglow predictions for the synchrotron maser model, as prescribed by \protect\cite{Margalit_2020} and \protect\cite{Cooper_2022}, using X-ray/radio flux ratios of $10^5$ and $10^8$. For this model, we used the luminosity value for FRB\,20230808F that was calculated using the burst bandwidth. The upper flux limits from our earliest post-FRB MeerLICHT observations are represented by the inverted triangles.}
    \label{fig:afterglow}
\end{figure}

Following the afterglow prescription outlined in \cite{Metzger_2019}, \cite{Margalit_2020} and \cite{Cooper_2022}, we used the observed properties of FRB\,20230808F to make a prediction for the optical afterglow expected within the maser shock framework at MeerLICHT's $q$-band frequency. The predicted afterglow, along with our MeerLICHT $q$-band flux upper limits obtained within the first 20 minutes post-burst, is shown in Figure~\ref{fig:afterglow}. We assume an unseen early X-ray afterglow with a given X-ray/radio flux luminosity, in order to normalise the rest of the afterglow, using values of $L_{\mathrm{X-ray}}/L_{\mathrm{radio}}=10^5$ and $L_{\mathrm{X-ray}}/L_{\mathrm{radio}}=10^8$. Our upper limits in this case, at such a large cosmological distance, are not constraining for the maser shock model, although they would be constraining for FRBs at lower redshifts, or for deeper optical flux limits.

\subsection{Host galaxy and redshift}
From our observations of the FRB region using SALT, we were able to obtain a spectroscopic redshift of $z_{\mathrm{spec}}=0.3472\pm0.0002$, and we took this to be the redshift of both galaxies and thus of FRB\,20230808F. We conclude that the two potential host galaxies are a close pair, and possibly interacting. Furthermore, we conclude that either the host galaxy, a foreground galaxy, or the FRB environment contribute significantly to the DM, scattering timescale, and rotation measure of the burst.

\section{Summary and conclusions}
We obtained an almost simultaneous optical limit associated with the newly discovered FRB\,20230808F. We set a flux limit of $f_{\mathrm{AB,}q}<11.7\,\micro \mathrm{Jy}$ for a 60\,s exposure starting $\sim$3.4\,s after the burst, which translates to a luminosity limit of $\nu L_{\nu}\sim1.3\times10^{43}\,\mathrm{erg\,s^{-1}}$. This is deeper than most other luminosity limits for optical emission associated with extragalactic FRBs, surpassed only by limits for repeaters obtained by \cite{Hiramatsu_2023}, \cite{kilpatrick_2023} and \cite{trudu}. Our limits were not constraining for the synchrotron maser model, but would be for an FRB at a lower redshift, or if our flux limits were deeper. The FRB was localised to a pair of close, possibly interacting galaxies at a redshift $z_{\mathrm{spec}}=0.3472\pm0.0002$, and the scattering, DM and RM were all found to have significant extragalactic contributions, from either the FRB environment, host galaxy, or a foreground galaxy. We have significantly improved upon previous ratios of $F_{\mathrm{opt}}/F_{\mathrm{radio}}$ for non-repeating FRBs with our $F_{\mathrm{opt}}/F_{\mathrm{radio}}\lesssim0.023$, when the optical fluence is calculated on a timescale equal to the delay between the FRB arrival time and the start of the optical exposure. Our delay of $\sim$3.4\,s is the shortest delay for an optical observation of a non-repeating FRB. We have demonstrated the ability of MeerLICHT's observing strategy when it comes to achieving almost simultaneous observations of FRBs, which has the potential to yield strictly simultaneous optical observations, even for non-repeating FRBs, in the future. With this setup, there are further opportunities for simultaneous MeerKAT-MeerLICHT observations of FRBs, and this should allow us to set more constraining limits on any optical emission associated with these enigmatic bursts.

\section*{Acknowledgements}
We acknowledge the use of the ilifu cloud computing facility – www.ilifu.ac.za, a partnership between the University of Cape Town, the University of the Western Cape, Stellenbosch University, Sol Plaatje University and the Cape Peninsula University of Technology. The ilifu facility is supported by contributions from the Inter-University Institute for Data Intensive Astronomy (IDIA – a partnership between the University of Cape Town, the University of Pretoria and the University of the Western Cape), the Computational Biology division at UCT and the Data Intensive Research Initiative of South Africa (DIRISA).

MeerLICHT is designed, built and operated by a consortium consisting of Radboud University, the University of Cape Town, the South African Astronomical Observatory (SAAO), the University of Oxford, the University of Manchester and the University of Amsterdam, in association with, and partly supported by, the South African Radio Astronomy Observatory (SARAO), the European Research Council, the Netherlands Research School for Astronomy (NOVA) and the Netherlands Organisation for Scientific Research (NWO).

The MeerKAT telescope is operated by the South African Radio
Astronomy Observatory, which is a facility of the National Research Foundation, an agency of the Department of Science and Innovation. SARAO acknowledges the ongoing advice and calibration of GPS systems by the National Metrology Institute of South Africa (NMISA) and the time space reference systems department of the Paris Observatory.

The MeerTRAP collaboration would like to thank the MeerKAT Large Survey Project teams for allowing MeerTRAP to observe commensally. The MeerTRAP collaboration acknowledges funding from the European Research Council under the European Union’s Horizon 2020 research and innovation programme (grant agreement No 694745).

We acknowledge the usage of TRAPUM infrastructure funded and installed by the Max-Planck-Institut für Radioastronomie and the Max-Planck-Gesellschaft.

This project used public archival data from the Dark Energy Survey (DES). Funding for the DES Projects has been provided by the U.S. Department of Energy, the U.S. National Science Foundation, the Ministry of Science and Education of Spain, the Science and Technology Facilities Council of the United Kingdom, the Higher Education Funding Council for England, the National Center for Supercomputing Applications at the University of Illinois at Urbana–Champaign, the Kavli Institute of Cosmological Physics at the University of Chicago, the Center for Cosmology and Astro-Particle Physics at the Ohio State University, the Mitchell Institute for Fundamental Physics and Astronomy at Texas A\&M University, Financiadora de Estudos e Projetos, Fundação Carlos Chagas Filho de Amparo à Pesquisa do Estado do Rio de Janeiro, Conselho Nacional de Desenvolvimento Científico e Tecnológico and the Ministério da Ciência, Tecnologia e Inovação, the Deutsche Forschungsgemeinschaft and the Collaborating Institutions in the Dark Energy Survey.

The Legacy Surveys consist of three individual and complementary projects: the Dark Energy Camera Legacy Survey (DECaLS; Proposal ID \#2014B-0404; PIs: David Schlegel and Arjun Dey), the Beijing-Arizona Sky Survey (BASS; NOAO Prop. ID \#2015A-0801; PIs: Zhou Xu and Xiaohui Fan), and the Mayall z-band Legacy Survey (MzLS; Prop. ID \#2016A-0453; PI: Arjun Dey). DECaLS, BASS and MzLS together include data obtained, respectively, at the Blanco telescope, Cerro Tololo Inter-American Observatory, NSF’s NOIRLab; the Bok telescope, Steward Observatory, University of Arizona; and the Mayall telescope, Kitt Peak National Observatory, NOIRLab. Pipeline processing and analyses of the data were supported by NOIRLab and the Lawrence Berkeley National Laboratory (LBNL). The Legacy Surveys project is honored to be permitted to conduct astronomical research on Iolkam Du’ag (Kitt Peak), a mountain with particular significance to the Tohono O’odham Nation.

IPM acknowledges funding from an NWO Rubicon Fellowship, project number 019.221EN.019.

KYH acknowledges the financial support provided by the South African Radio Astronomy Observatory (SARAO).

PJG is partly supported by NRF SARChI grant 111692.

\section*{Data Availability}
The data will be made available upon reasonable request to the authors.

\onecolumn
\begin{longtable}[c]{ c  c  c !{\vrule width 1pt} c  c  c  c}
 \caption{Upper limits on optical emission for FRB\,20230808F, corrected for MW dust extinction.}
 \label{tab:Results_Extinct}\\

 \hline
 \makecell{\textbf{Filter}\\\textbf{}} & \makecell{\textbf{Upper limit or detection}\\\textbf{magnitude (AB)}} & \makecell{\textbf{Upper limit or}\\\textbf{detection flux} ($\mathrm{\micro Jy}$)} &  & \makecell{\textbf{Filter}\\\textbf{}} & \makecell{\textbf{Upper limit or detection}\\\textbf{magnitude (AB)}} & \makecell{\textbf{Upper limit or}\\\textbf{detection flux} ($\mathrm{\micro Jy}$)} \\
 \hline
 \endfirsthead

 % \hline
 % \multicolumn{2}{|c|}{Continuation of Table \ref{long}}\\
 \hline
 \makecell{\textbf{Filter}\\\textbf{}} & \makecell{\textbf{Upper limit or detection}\\\textbf{magnitude (AB)}} & \makecell{\textbf{Upper limit or}\\\textbf{detection flux} ($\mathrm{\micro Jy}$)} &  & \makecell{\textbf{Filter}\\\textbf{}} & \makecell{\textbf{Upper limit or detection}\\\textbf{magnitude (AB)}} & \makecell{\textbf{Upper limit or}\\\textbf{detection flux} ($\mathrm{\micro Jy}$)}\\
 \hline
 \endhead

 \hline
 \endfoot

 \hline
 \multicolumn{3}{ c }{}\\
 \hline\hline
 \endlastfoot

\bottomrule
&\textbf{1 min before}& & & &\textbf{1 min after} &\\
\specialrule{0.8pt}{0pt}{0pt}
\textit{u}               &  19.7  & 46.9 & & \textit{q} & 21.2  & 11.7 \\  
\toprule
\bottomrule
&\textbf{5 min before}& & & &\textbf{5 min after} &\\
\specialrule{0.8pt}{0pt}{0pt}
\textit{u}               &  19.7  & 46.9 & & \textit{q} & 21.6  & 8.2 \\  
\textit{q}               &  21.7  & 7.5 & & \textit{r} & 20.6  & 20.0\\  
\textit{i}               &  20.3  & 28.5 & & \textit{z} & 19.2  & 74.7\\ 
\toprule
\bottomrule
&\textbf{15 min before}& & & &\textbf{15 min after} &\\
\specialrule{0.8pt}{0pt}{0pt}
\textit{u}               &  19.7  & 46.9 & & \textit{u} &  19.5  & 57.0 \\ 
\textit{g}               &  20.9  & 16.5 & & \textit{g}  &  20.7  & 19.4 \\ 
\textit{q}               &  22.2  & 4.7 & & \textit{q}  &  22.0  & 5.7 \\ 
\textit{r}               &  20.7  & 19.1 & & \textit{r} &  20.6  & 20.0 \\ 
\textit{i}               &  20.3  & 28.5 & & \textit{i} &  20.0  & 36.4 \\ 
\textit{z}               &  19.3  & 68.8 & & \textit{z} &  19.2  & 74.7 \\ 
\toprule
\bottomrule
&\textbf{20 min before}& & & &\textbf{20 min after} &\\
\specialrule{0.8pt}{0pt}{0pt}
\textit{u}               &  20.1  & 32.9 & & \textit{u} &  19.5  & 57.0 \\ 
\textit{g}               &  20.9  & 16.5 & & \textit{g}  &  20.7  & 19.4 \\ 
\textit{q}               &  22.4  & 4.0 & & \textit{q}  &  22.1  & 5.2 \\ 
\textit{r}               &  20.7  & 19.1 & & \textit{r} &  20.9  & 15.3 \\ 
\textit{i}               &  20.6  & 20.0 & & \textit{i} &  20.0  & 36.4 \\ 
\textit{z}               &  19.3  & 68.8 & & \textit{z} &  19.2  & 74.7 \\
\toprule
\bottomrule
&\textbf{2023-08-01}& & & &\textbf{2023-08-02} &\\
\specialrule{0.8pt}{0pt}{0pt}
\textit{u}               &  20.4  & 26.1 & & \textit{u} &  19.8  & 42.2  \\
\textit{g}               &  21.2  & 12.1 & & \textit{g} &  20.7  & 18.4 \\
\textit{q}               &  22.5  & 3.7 & & \textit{q}  &  22.1  & 5.4 \\
\textit{r}               &  21.2  & 12.0 & & \textit{r} &  20.8  & 17.3 \\
\textit{i}               &  21.1  & 13.1 & & \textit{i} &  20.5  & 22.1 \\
\textit{z}               &  20.3  & 28.5 & & \textit{z} &  19.6  & 50.2 \\   
\toprule
\bottomrule
&\textbf{2023-09-06}& & & &\textbf{2023-10-01} & \\
\specialrule{0.8pt}{0pt}{0pt}
\textit{u}               &  22.0  & 6.0 & & \textit{u} &  19.0  & 95.5 \\
\textit{g}               &  22.5  & 3.7 & & \textit{q} &  20.5  & 22.5 \\
\textit{q}               &  $22.3\pm0.2$  & $4.3\pm0.6$ & & i &  19.9  & 39.9 \\
\textit{i}               &  $21.7\pm0.2$  & $7.5\pm1.6$  \\
\textit{z}               &  21.3  & 10.7  \\  
\toprule
\bottomrule
&\textbf{2023-10-02}& & & &\textbf{2023-10-03} \\
\specialrule{0.8pt}{0pt}{0pt}
\textit{u}               &  19.1  & 85.9 & & \textit{u} &  19.5  & 57.5 \\
\textit{q}               &  20.6  & 20.1 & & \textit{q} &  21.4 & 10.1 \\
\textit{i}               &  19.8  & 44.7 & & \textit{i} &  20.1  & 32.0 \\
\toprule
\bottomrule
&\textbf{2023-10-04}& & & &\textbf{2023-10-17}\\
\specialrule{0.8pt}{0pt}{0pt}
\textit{u}               &  18.5  & 151.0 & & \textit{u} &  20.6  & 21.8 \\
\textit{q}               &  20.4  & 24.7 & & \textit{q}  &  $22.2\pm0.3$  & $4.7\pm1.3$ \\
\textit{i}               &  19.1  & 85.8 & & \textit{i}  &  21.2  & 12.4 \\  
\toprule
\bottomrule
&\textbf{2023-10-18}& & & &\textbf{2023-10-19} \\
\specialrule{0.8pt}{0pt}{0pt}
\textit{u}               &  20.0  & 37.4 & & \textit{u} &  21.9  & 6.0 \\
\textit{q}               &  21.8  & 6.9  & & \textit{q} &  $22.3\pm0.1$  & $4.3\pm0.6$ \\
\textit{i}               &  20.2  & 29.7 & & \textit{i} &  $21.8\pm0.3$  & $6.7\pm1.6$ \\   
\toprule
\bottomrule
&\textbf{2023-10-25}& & & &\textbf{2023-10-26}\\
\specialrule{0.8pt}{0pt}{0pt}
\textit{u}               &  21.1  & 13.8 & & \textit{u} &  20.1  & 33.3 \\
\textit{q}               &  22.5  & 3.8  & & \textit{q} &  21.5  & 8.9 \\
\textit{i}               &  21.6  & 8.5 & & \textit{i}  &  20.8  & 17.7 \\  
\toprule
\bottomrule
&\textbf{2023-10-29}& & & &\textbf{2023-10-30}\\
\specialrule{0.8pt}{0pt}{0pt}
\textit{q}               &  20.0  & 37.0 & & \textit{u} &  20.0  & 37.5 \\
 & & & & \textit{q} &  21.3  & 10.7  \\
 & & & & \textit{i} &  20.6  & 21.0  \\ 
\toprule
\bottomrule
&\textbf{2023-10-31}& & & &\textbf{2023-11-01} \\
\specialrule{0.8pt}{0pt}{0pt}
\textit{u}               &  20.0  & 37.1 & & \textit{u} &  19.3  & 71.2 \\
\textit{q}               &  21.4  & 9.7 & & \textit{q} &  21.4  & 9.9 \\
\textit{i}               &  19.8  & 44.7  \\  
\toprule
\bottomrule
&\textbf{2023-11-02}& & & &\textbf{2023-11-04} \\
\specialrule{0.8pt}{0pt}{0pt}
\textit{u}               &  19.2  & 78.1 & & \textit{u} &  21.0  & 14.0 \\
\textit{q}               &  21.0  & 14.7 & & \textit{q} &  $22.3\pm0.2$  & $4.2\pm0.9$ \\
\textit{i}               &  19.5  & 59.4 & & \textit{i} &  21.5  & 8.9 \\   
\toprule
\bottomrule
&\textbf{2023-11-07}& & & &\textbf{3 months post-FRB}\\
\specialrule{0.8pt}{0pt}{0pt}
\textit{u}               &  20.7  & 19.9 & & \textit{u} &  22.3  & 4.3 \\
\textit{q}               &  $21.9\pm0.2$  & $6.5\pm1.3$ & & \textit{g} &  22.4  & 3.8 \\
\textit{i}               &  21.0  & 14.5 & & \textit{q} &  $23.8\pm0.1$  & $1.1\pm0.1$ \\
& & & & \textit{r} &  20.9  & 15.3  \\
& & & & \textit{i} &  $22.0\pm0.2$  & $5.7\pm1.1$  \\
& & & & \textit{z} &  21.3  & 10.7  
 \end{longtable}
 \twocolumn

\bibliographystyle{mnras}
\bibliography{bibliography}

\begin{thebibliography}{}
\makeatletter
\relax
\def\mn@urlcharsother{\let\do\@makeother \do\$\do\&\do\#\do\^\do\_\do\%\do\~}
\def\mn@doi{\begingroup\mn@urlcharsother \@ifnextchar [ {\mn@doi@} {\mn@doi@[]}}
\def\mn@doi@[#1]#2{\def\@tempa{#1}\ifx\@tempa\@empty \href {http://dx.doi.org/#2} {doi:#2}\else \href {http://dx.doi.org/#2} {#1}\fi \endgroup}
\def\mn@eprint#1#2{\mn@eprint@#1:#2::\@nil}
\def\mn@eprint@arXiv#1{\href {http://arxiv.org/abs/#1} {{\tt arXiv:#1}}}
\def\mn@eprint@dblp#1{\href {http://dblp.uni-trier.de/rec/bibtex/#1.xml} {dblp:#1}}
\def\mn@eprint@#1:#2:#3:#4\@nil{\def\@tempa {#1}\def\@tempb {#2}\def\@tempc {#3}\ifx \@tempc \@empty \let \@tempc \@tempb \let \@tempb \@tempa \fi \ifx \@tempb \@empty \def\@tempb {arXiv}\fi \@ifundefined {mn@eprint@\@tempb}{\@tempb:\@tempc}{\expandafter \expandafter \csname mn@eprint@\@tempb\endcsname \expandafter{\@tempc}}}

\bibitem[\protect\citeauthoryear{Aartsen et~al.,}{Aartsen et~al.}{2018}]{Aartsen_2018}
Aartsen M.~G.,  et~al., 2018, \mn@doi [The Astrophysical Journal] {10.3847/1538-4357/aab4f8}, 857, 117

\bibitem[\protect\citeauthoryear{{Abbott} et~al.,}{{Abbott} et~al.}{2021}]{DES_dr2}
{Abbott} T.~M.~C.,  et~al., 2021, \mn@doi [\apjs] {10.3847/1538-4365/ac00b3}, \href {https://ui.adsabs.harvard.edu/abs/2021ApJS..255...20A} {255, 20}

\bibitem[\protect\citeauthoryear{Acciari et~al.,}{Acciari et~al.}{2018}]{Acciari}
Acciari V.~A.,  et~al., 2018, \mn@doi [Monthly Notices of the Royal Astronomical Society] {10.1093/mnras/sty2422}, 481, 2479–2486

\bibitem[\protect\citeauthoryear{Adámek \& Armour}{Adámek \& Armour}{2016}]{adámek2016realtime}
Adámek K.,  Armour W.,  2016, A Real-time Single Pulse Detection Algorithm for GPUs (\mn@eprint {arXiv} {1611.09704})

\bibitem[\protect\citeauthoryear{Adámek \& Armour}{Adámek \& Armour}{2018}]{adámek2018gpu}
Adámek K.,  Armour W.,  2018, A GPU implementation of the harmonic sum algorithm (\mn@eprint {arXiv} {1812.02647})

\bibitem[\protect\citeauthoryear{Adámek, Dimoudi, Giles  \& Armour}{Adámek et~al.}{2017}]{adámek2017improved}
Adámek K.,  Dimoudi S.,  Giles M.,   Armour W.,  2017, Improved Acceleration of the GPU Fourier Domain Acceleration Search Algorithm (\mn@eprint {arXiv} {1711.10855})

\bibitem[\protect\citeauthoryear{Aggarwal}{Aggarwal}{2021}]{Aggarwal_2021}
Aggarwal K.,  2021, \mn@doi [The Astrophysical Journal Letters] {10.3847/2041-8213/ac2a3a}, 920, L18

\bibitem[\protect\citeauthoryear{{Aggarwal}, {Budav{\'a}ri}, {Deller}, {Eftekhari}, {James}, {Prochaska}  \& {Tendulkar}}{{Aggarwal} et~al.}{2021}]{path}
{Aggarwal} K.,  {Budav{\'a}ri} T.,  {Deller} A.~T.,  {Eftekhari} T.,  {James} C.~W.,  {Prochaska} J.~X.,   {Tendulkar} S.~P.,  2021, \mn@doi [\apj] {10.3847/1538-4357/abe8d2}, \href {https://ui.adsabs.harvard.edu/abs/2021ApJ...911...95A} {911, 95}

\bibitem[\protect\citeauthoryear{{Ambrosino}, {Cretaro}, {Meddi}, {Rossi}, {Sclavi}  \& {Bruni}}{{Ambrosino} et~al.}{2016}]{Ambrosiono_2016}
{Ambrosino} F.,  {Cretaro} P.,  {Meddi} F.,  {Rossi} C.,  {Sclavi} S.,   {Bruni} I.,  2016, \mn@doi [Journal of Astronomical Instrumentation] {10.1142/S2251171716500057}, \href {https://ui.adsabs.harvard.edu/abs/2016JAI.....550005A} {5, 1650005}

\bibitem[\protect\citeauthoryear{Bailes et~al.,}{Bailes et~al.}{2021}]{Bailes_2021}
Bailes M.,  et~al., 2021, \mn@doi [Monthly Notices of the Royal Astronomical Society] {10.1093/mnras/stab749}, 503, 5367–5384

\bibitem[\protect\citeauthoryear{{Barbieri} et~al.,}{{Barbieri} et~al.}{2009}]{Barbieri_2009}
{Barbieri} C.,  et~al., 2009, \mn@doi [Journal of Modern Optics] {10.1080/09500340802450565}, \href {https://ui.adsabs.harvard.edu/abs/2009JMOp...56..261B} {56, 261}

\bibitem[\protect\citeauthoryear{Barr}{Barr}{2017}]{Barr_2017}
Barr E.~D.,  2017, \mn@doi [Proceedings of the International Astronomical Union] {10.1017/S1743921317009036}, 13, 175–178

\bibitem[\protect\citeauthoryear{{Beloborodov}}{{Beloborodov}}{2017}]{Beloborodov_2017}
{Beloborodov} A.~M.,  2017, \mn@doi [\apjl] {10.3847/2041-8213/aa78f3}, \href {https://ui.adsabs.harvard.edu/abs/2017ApJ...843L..26B} {843, L26}

\bibitem[\protect\citeauthoryear{{Beloborodov}}{{Beloborodov}}{2020}]{Beloborodov_2020}
{Beloborodov} A.~M.,  2020, \mn@doi [\apj] {10.3847/1538-4357/ab83eb}, \href {https://ui.adsabs.harvard.edu/abs/2020ApJ...896..142B} {896, 142}

\bibitem[\protect\citeauthoryear{Beroiz, Cabral  \& Sanchez}{Beroiz et~al.}{2020}]{astroalign}
Beroiz M.,  Cabral J.,   Sanchez B.,  2020, \mn@doi [Astronomy and Computing] {https://doi.org/10.1016/j.ascom.2020.100384}, 32, 100384

\bibitem[\protect\citeauthoryear{Bezuidenhout et~al.,}{Bezuidenhout et~al.}{2022}]{Bezuidenhout_2022}
Bezuidenhout M.~C.,  et~al., 2022, \mn@doi [Monthly Notices of the Royal Astronomical Society] {10.1093/mnras/stac579}, 512, 1483

\bibitem[\protect\citeauthoryear{{Bhardwaj} et~al.,}{{Bhardwaj} et~al.}{2021}]{Bhardwaj}
{Bhardwaj} M.,  et~al., 2021, \mn@doi [\apjl] {10.3847/2041-8213/abeaa6}, \href {https://ui.adsabs.harvard.edu/abs/2021ApJ...910L..18B} {910, L18}

\bibitem[\protect\citeauthoryear{{Bochenek}, {Ravi}, {Belov}, {Hallinan}, {Kocz}, {Kulkarni}  \& {McKenna}}{{Bochenek} et~al.}{2020}]{bochenek_2020}
{Bochenek} C.~D.,  {Ravi} V.,  {Belov} K.~V.,  {Hallinan} G.,  {Kocz} J.,  {Kulkarni} S.~R.,   {McKenna} D.~L.,  2020, \mn@doi [\nat] {10.1038/s41586-020-2872-x}, \href {https://ui.adsabs.harvard.edu/abs/2020Natur.587...59B} {587, 59}

\bibitem[\protect\citeauthoryear{Brentjens \& de Bruyn}{Brentjens \& de~Bruyn}{2005}]{Brentjens_2005}
Brentjens M.~A.,  de Bruyn A.~G.,  2005, \mn@doi [Astronomy &amp; Astrophysics] {10.1051/0004-6361:20052990}, 441, 1217–1228

\bibitem[\protect\citeauthoryear{{Brown} et~al.,}{{Brown} et~al.}{2013a}]{LCO}
{Brown} T.~M.,  et~al., 2013a, \mn@doi [\pasp] {10.1086/673168}, \href {https://ui.adsabs.harvard.edu/abs/2013PASP..125.1031B} {125, 1031}

\bibitem[\protect\citeauthoryear{{Brown} et~al.,}{{Brown} et~al.}{2013b}]{Brown_2013}
{Brown} T.~M.,  et~al., 2013b, \mn@doi [\pasp] {10.1086/673168}, \href {https://ui.adsabs.harvard.edu/abs/2013PASP..125.1031B} {125, 1031}

\bibitem[\protect\citeauthoryear{Brown et~al.,}{Brown et~al.}{2018}]{Gaia_2018}
Brown A. G.~A.,  et~al., 2018, \mn@doi [Astronomy &amp; Astrophysics] {10.1051/0004-6361/201833051}, 616, A1

\bibitem[\protect\citeauthoryear{{Buckley}, {Swart}  \& {Meiring}}{{Buckley} et~al.}{2006}]{SALT_2006}
{Buckley} D. A.~H.,  {Swart} G.~P.,   {Meiring} J.~G.,  2006, in {Stepp} L.~M.,  ed.,  Society of Photo-Optical Instrumentation Engineers (SPIE) Conference Series Vol. 6267, Ground-based and Airborne Telescopes. p. 62670Z, \mn@doi{10.1117/12.673750}

\bibitem[\protect\citeauthoryear{{Burgh}, {Nordsieck}, {Kobulnicky}, {Williams}, {O'Donoghue}, {Smith}  \& {Percival}}{{Burgh} et~al.}{2003}]{rss}
{Burgh} E.~B.,  {Nordsieck} K.~H.,  {Kobulnicky} H.~A.,  {Williams} T.~B.,  {O'Donoghue} D.,  {Smith} M.~P.,   {Percival} J.~W.,  2003, in {Iye} M.,  {Moorwood} A. F.~M.,  eds,  Society of Photo-Optical Instrumentation Engineers (SPIE) Conference Series Vol. 4841, Instrument Design and Performance for Optical/Infrared Ground-based Telescopes. pp 1463--1471, \mn@doi{10.1117/12.460312}

\bibitem[\protect\citeauthoryear{Burn}{Burn}{1966}]{Burn_1966}
Burn B.~J.,  1966, \mn@doi [Monthly Notices of the Royal Astronomical Society] {10.1093/mnras/133.1.67}, 133, 67

\bibitem[\protect\citeauthoryear{{CHIME/FRB Collaboration} et~al.,}{{CHIME/FRB Collaboration} et~al.}{2019}]{CHIME_2019}
{CHIME/FRB Collaboration} et~al., 2019, \mn@doi [\apjl] {10.3847/2041-8213/ab4a80}, \href {https://ui.adsabs.harvard.edu/abs/2019ApJ...885L..24C} {885, L24}

\bibitem[\protect\citeauthoryear{{CHIME/FRB Collaboration} et~al.,}{{CHIME/FRB Collaboration} et~al.}{2020}]{2020Natur.587...54C}
{CHIME/FRB Collaboration} et~al., 2020, \mn@doi [\nat] {10.1038/s41586-020-2863-y}, \href {https://ui.adsabs.harvard.edu/abs/2020Natur.587...54C} {587, 54}

\bibitem[\protect\citeauthoryear{{CHIME/FRB Collaboration} et~al.,}{{CHIME/FRB Collaboration} et~al.}{2023}]{Chime_2023}
{CHIME/FRB Collaboration} et~al., 2023, \mn@doi [\apj] {10.3847/1538-4357/acc6c1}, \href {https://ui.adsabs.harvard.edu/abs/2023ApJ...947...83C} {947, 83}

\bibitem[\protect\citeauthoryear{Caleb et~al.,}{Caleb et~al.}{2020}]{Caleb_2020}
Caleb M.,  et~al., 2020, \mn@doi [Monthly Notices of the Royal Astronomical Society] {10.1093/mnras/staa1791}, 496, 4565

\bibitem[\protect\citeauthoryear{Callister, Kanner  \& Weinstein}{Callister et~al.}{2016}]{Callister_2016}
Callister T.,  Kanner J.,   Weinstein A.,  2016, \mn@doi [The Astrophysical Journal Letters] {10.3847/2041-8205/825/1/L12}, 825, L12

\bibitem[\protect\citeauthoryear{Chen, Barr, Karuppusamy, Kramer  \& Stappers}{Chen et~al.}{2021}]{Chen_2021}
Chen W.,  Barr E.,  Karuppusamy R.,  Kramer M.,   Stappers B.,  2021, \mn@doi [Journal of Astronomical Instrumentation] {10.1142/s2251171721500136}, 10

\bibitem[\protect\citeauthoryear{Cho et~al.,}{Cho et~al.}{2020}]{Cho_2020}
Cho H.,  et~al., 2020, \mn@doi [The Astrophysical Journal Letters] {10.3847/2041-8213/ab7824}, 891, L38

\bibitem[\protect\citeauthoryear{{Connor} et~al.,}{{Connor} et~al.}{2024}]{Connor_2024}
{Connor} L.,  et~al., 2024, \mn@doi [arXiv e-prints] {10.48550/arXiv.2409.16952}, \href {https://ui.adsabs.harvard.edu/abs/2024arXiv240916952C} {p. arXiv:2409.16952}

\bibitem[\protect\citeauthoryear{{Cooper} \& {Wijers}}{{Cooper} \& {Wijers}}{2021}]{Cooper_2021}
{Cooper} A.~J.,  {Wijers} R.~A.~M.~J.,  2021, \mn@doi [Monthly Notices of the Royal Astronomical Society] {10.1093/mnrasl/slab099}, \href {https://ui.adsabs.harvard.edu/abs/2021MNRAS.508L..32C} {508, L32}

\bibitem[\protect\citeauthoryear{Cooper et~al.,}{Cooper et~al.}{2022}]{Cooper_2022}
Cooper A.~J.,  et~al., 2022, \mn@doi [Monthly Notices of the Royal Astronomical Society] {10.1093/mnras/stac2951}, 517, 5483–5495

\bibitem[\protect\citeauthoryear{Cordes \& Lazio}{Cordes \& Lazio}{2003}]{Cordes_2003}
Cordes J.~M.,  Lazio T. J.~W.,  2003, NE2001.I. A New Model for the Galactic Distribution of Free Electrons and its Fluctuations (\mn@eprint {arXiv} {astro-ph/0207156}), \url {https://arxiv.org/abs/astro-ph/0207156}

\bibitem[\protect\citeauthoryear{Crawford, Hisano, Golden, Kikunaga, Laity  \& Zoeller}{Crawford et~al.}{2022}]{Crawford_2022}
Crawford F.,  Hisano S.,  Golden M.,  Kikunaga T.,  Laity A.,   Zoeller D.,  2022, \mn@doi [Monthly Notices of the Royal Astronomical Society] {10.1093/mnras/stac2101}, 515, 3698

\bibitem[\protect\citeauthoryear{{Cummings}, {Barthelmy}, {Chester}  \& {Page}}{{Cummings} et~al.}{2014}]{Cummings_2014}
{Cummings} J.~R.,  {Barthelmy} S.~D.,  {Chester} M.~M.,   {Page} K.~L.,  2014, The Astronomer's Telegram, \href {https://ui.adsabs.harvard.edu/abs/2014ATel.6294....1C} {6294, 1}

\bibitem[\protect\citeauthoryear{{DESI Collaboration} et~al.,}{{DESI Collaboration} et~al.}{2016}]{desi_2016}
{DESI Collaboration} et~al., 2016, The DESI Experiment Part I: Science,Targeting, and Survey Design (\mn@eprint {arXiv} {1611.00036}), \url {https://arxiv.org/abs/1611.00036}

\bibitem[\protect\citeauthoryear{Day et~al.,}{Day et~al.}{2020}]{Day_2020}
Day C.~K.,  et~al., 2020, \mn@doi [Monthly Notices of the Royal Astronomical Society] {10.1093/mnras/staa2138}, 497, 3335–3350

\bibitem[\protect\citeauthoryear{{Dewey}, {Taylor}, {Weisberg}  \& {Stokes}}{{Dewey} et~al.}{1985}]{Dewey_1985}
{Dewey} R.~J.,  {Taylor} J.~H.,  {Weisberg} J.~M.,   {Stokes} G.~H.,  1985, \mn@doi [\apjl] {10.1086/184502}, \href {https://ui.adsabs.harvard.edu/abs/1985ApJ...294L..25D} {294, L25}

\bibitem[\protect\citeauthoryear{Dhillon et~al.,}{Dhillon et~al.}{2014}]{Ultraspec}
Dhillon V.~S.,  et~al., 2014, \mn@doi [Monthly Notices of the Royal Astronomical Society] {10.1093/mnras/stu1660}, 444, 4009

\bibitem[\protect\citeauthoryear{Dimoudi \& Armour}{Dimoudi \& Armour}{2015}]{dimoudi2015pulsar}
Dimoudi S.,  Armour W.,  2015, Pulsar Acceleration Searches on the GPU for the Square Kilometre Array (\mn@eprint {arXiv} {1511.07343})

\bibitem[\protect\citeauthoryear{Dimoudi, Adamek, Thiagaraj, Ransom, Karastergiou  \& Armour}{Dimoudi et~al.}{2018}]{Dimoudi_2018}
Dimoudi S.,  Adamek K.,  Thiagaraj P.,  Ransom S.~M.,  Karastergiou A.,   Armour W.,  2018, \mn@doi [The Astrophysical Journal Supplement Series] {10.3847/1538-4365/aabe88}, 239, 28

\bibitem[\protect\citeauthoryear{Driessen et~al.,}{Driessen et~al.}{2023}]{driessen2023frb}
Driessen L.~N.,  et~al., 2023, FRB 20210405I: a nearby Fast Radio Burst localised to sub-arcsecond precision with MeerKAT (\mn@eprint {arXiv} {2302.09787})

\bibitem[\protect\citeauthoryear{Duchesne et~al.,}{Duchesne et~al.}{2023}]{Duchesne_2023}
Duchesne S.~W.,  et~al., 2023, \mn@doi [Publications of the Astronomical Society of Australia] {10.1017/pasa.2023.31}, 40

\bibitem[\protect\citeauthoryear{{Eftekhari}, {Berger}, {Williams}  \& {Blanchard}}{{Eftekhari} et~al.}{2018}]{Eftekhari}
{Eftekhari} T.,  {Berger} E.,  {Williams} P.~K.~G.,   {Blanchard} P.~K.,  2018, \mn@doi [\apj] {10.3847/1538-4357/aac270}, \href {https://ui.adsabs.harvard.edu/abs/2018ApJ...860...73E} {860, 73}

\bibitem[\protect\citeauthoryear{{Fabricant} et~al.,}{{Fabricant} et~al.}{2019}]{Binospec}
{Fabricant} D.,  et~al., 2019, \mn@doi [\pasp] {10.1088/1538-3873/ab1d78}, \href {https://ui.adsabs.harvard.edu/abs/2019PASP..131g5004F} {131, 075004}

\bibitem[\protect\citeauthoryear{Falcke \& Rezzolla}{Falcke \& Rezzolla}{2014}]{Falcke_2014}
Falcke H.,  Rezzolla L.,  2014, \mn@doi [Astronomy &amp; Astrophysics] {10.1051/0004-6361/201321996}, 562, A137

\bibitem[\protect\citeauthoryear{{Fitzpatrick}}{{Fitzpatrick}}{1999}]{Fitzpatrick}
{Fitzpatrick} E.~L.,  1999, \mn@doi [\pasp] {10.1086/316293}, \href {https://ui.adsabs.harvard.edu/abs/1999PASP..111...63F} {111, 63}

\bibitem[\protect\citeauthoryear{{Fonseca} et~al.,}{{Fonseca} et~al.}{2020}]{Fonseca_2020}
{Fonseca} E.,  et~al., 2020, \mn@doi [\apjl] {10.3847/2041-8213/ab7208}, \href {https://ui.adsabs.harvard.edu/abs/2020ApJ...891L...6F} {891, L6}

\bibitem[\protect\citeauthoryear{{Ghedina} et~al.,}{{Ghedina} et~al.}{2018}]{Ghedina_2018}
{Ghedina} A.,  et~al., 2018, in {Evans} C.~J.,  {Simard} L.,   {Takami} H.,  eds,  Society of Photo-Optical Instrumentation Engineers (SPIE) Conference Series Vol. 10702, Ground-based and Airborne Instrumentation for Astronomy VII. p. 107025Q, \mn@doi{10.1117/12.2316348}

\bibitem[\protect\citeauthoryear{{Groot} et~al.,}{{Groot} et~al.}{2024}]{Groot_2024}
{Groot} P.~J.,  et~al., 2024, \mn@doi [arXiv e-prints] {10.48550/arXiv.2405.18923}, \href {https://ui.adsabs.harvard.edu/abs/2024arXiv240518923G} {p. arXiv:2405.18923}

\bibitem[\protect\citeauthoryear{{Gupta} et~al.,}{{Gupta} et~al.}{2017}]{Gupta_2017}
{Gupta} Y.,  et~al., 2017, \mn@doi [Current Science] {10.18520/cs/v113/i04/707-714}, \href {https://ui.adsabs.harvard.edu/abs/2017CSci..113..707G} {113, 707}

\bibitem[\protect\citeauthoryear{Hardy et~al.,}{Hardy et~al.}{2017}]{Hardy_2017}
Hardy L.~K.,  et~al., 2017, \mn@doi [Monthly Notices of the Royal Astronomical Society] {10.1093/mnras/stx2153}, 472, 2800–2807

\bibitem[\protect\citeauthoryear{Hessels et~al.,}{Hessels et~al.}{2019}]{Hessels_2019}
Hessels J. W.~T.,  et~al., 2019, \mn@doi [The Astrophysical Journal Letters] {10.3847/2041-8213/ab13ae}, 876, L23

\bibitem[\protect\citeauthoryear{Hiramatsu et~al.,}{Hiramatsu et~al.}{2023}]{Hiramatsu_2023}
Hiramatsu D.,  et~al., 2023, \mn@doi [The Astrophysical Journal Letters] {10.3847/2041-8213/acae98}, 947, L28

\bibitem[\protect\citeauthoryear{{Hutschenreuter} et~al.,}{{Hutschenreuter} et~al.}{2022}]{Hutschenreuter_2020}
{Hutschenreuter} S.,  et~al., 2022, \mn@doi [\aap] {10.1051/0004-6361/202140486}, \href {https://ui.adsabs.harvard.edu/abs/2022A&A...657A..43H} {657, A43}

\bibitem[\protect\citeauthoryear{Israel et~al.,}{Israel et~al.}{2016}]{Israel_2016}
Israel G.~L.,  et~al., 2016, \mn@doi [Monthly Notices of the Royal Astronomical Society] {10.1093/mnras/stw008}, 457, 3448

\bibitem[\protect\citeauthoryear{{Jankowski} et~al.,}{{Jankowski} et~al.}{2022}]{Jankowski_2022}
{Jankowski} F.,  et~al., 2022, in {Ruiz} J.~E.,  {Pierfedereci} F.,   {Teuben} P.,  eds,  Astronomical Society of the Pacific Conference Series Vol. 532, Astronomical Society of the Pacific Conference Series. p.~273 (\mn@eprint {arXiv} {2012.05173}), \mn@doi{10.48550/arXiv.2012.05173}

\bibitem[\protect\citeauthoryear{{Jankowski} et~al.,}{{Jankowski} et~al.}{2023}]{jankowski_flux}
{Jankowski} F.,  et~al., 2023, \mn@doi [\mnras] {10.1093/mnras/stad2041}, \href {https://ui.adsabs.harvard.edu/abs/2023MNRAS.524.4275J} {524, 4275}

\bibitem[\protect\citeauthoryear{{Jonas} \& {MeerKAT Team}}{{Jonas} \& {MeerKAT Team}}{2016}]{Jonas_2016}
{Jonas} J.,  {MeerKAT Team} 2016, in MeerKAT Science: On the Pathway to the SKA. p.~1, \mn@doi{10.22323/1.277.0001}

\bibitem[\protect\citeauthoryear{{Katz}}{{Katz}}{2016}]{Katz_2016}
{Katz} J.~I.,  2016, \mn@doi [\apj] {10.3847/0004-637X/826/2/226}, \href {https://ui.adsabs.harvard.edu/abs/2016ApJ...826..226K} {826, 226}

\bibitem[\protect\citeauthoryear{{Kilpatrick} et~al.,}{{Kilpatrick} et~al.}{2021}]{kilpatrick_2021}
{Kilpatrick} C.~D.,  et~al., 2021, \mn@doi [\apjl] {10.3847/2041-8213/abd560}, \href {https://ui.adsabs.harvard.edu/abs/2021ApJ...907L...3K} {907, L3}

\bibitem[\protect\citeauthoryear{{Kilpatrick} et~al.,}{{Kilpatrick} et~al.}{2024}]{kilpatrick_2023}
{Kilpatrick} C.~D.,  et~al., 2024, \mn@doi [\apj] {10.3847/1538-4357/ad2687}, \href {https://ui.adsabs.harvard.edu/abs/2024ApJ...964..121K} {964, 121}

\bibitem[\protect\citeauthoryear{{Kirsten} et~al.,}{{Kirsten} et~al.}{2022}]{Kirsten}
{Kirsten} F.,  et~al., 2022, \mn@doi [\nat] {10.1038/s41586-021-04354-w}, \href {https://ui.adsabs.harvard.edu/abs/2022Natur.602..585K} {602, 585}

\bibitem[\protect\citeauthoryear{{Kouwenhoven} \& {Vo{\^u}te}}{{Kouwenhoven} \& {Vo{\^u}te}}{2001}]{Kouwenhoven_2001}
{Kouwenhoven} M.~L.~A.,  {Vo{\^u}te} J.~L.~L.,  2001, \mn@doi [\aap] {10.1051/0004-6361:20011226}, \href {https://ui.adsabs.harvard.edu/abs/2001A&A...378..700K} {378, 700}

\bibitem[\protect\citeauthoryear{{Kumar} \& {Bo{\v{s}}njak}}{{Kumar} \& {Bo{\v{s}}njak}}{2020}]{Kumar_2020}
{Kumar} P.,  {Bo{\v{s}}njak} {\v{Z}}.,  2020, \mn@doi [\mnras] {10.1093/mnras/staa774}, \href {https://ui.adsabs.harvard.edu/abs/2020MNRAS.494.2385K} {494, 2385}

\bibitem[\protect\citeauthoryear{Kumar, Lu  \& Bhattacharya}{Kumar et~al.}{2017}]{Kumar_2017}
Kumar P.,  Lu W.,   Bhattacharya M.,  2017, \mn@doi [Monthly Notices of the Royal Astronomical Society] {10.1093/mnras/stx665}, 468, 2726

\bibitem[\protect\citeauthoryear{{Lien} et~al.,}{{Lien} et~al.}{2014}]{Lien_2014}
{Lien} A.~Y.,  et~al., 2014, GRB Coordinates Network, \href {https://ui.adsabs.harvard.edu/abs/2014GCN.16522....1L} {16522, 1}

\bibitem[\protect\citeauthoryear{Lorimer, Bailes, McLaughlin, Narkevic  \& Crawford}{Lorimer et~al.}{2007}]{Lorimer_2007}
Lorimer D.~R.,  Bailes M.,  McLaughlin M.~A.,  Narkevic D.~J.,   Crawford F.,  2007, \mn@doi [Science] {10.1126/science.1147532}, 318, 777–780

\bibitem[\protect\citeauthoryear{Luo et~al.,}{Luo et~al.}{2020}]{Luo_2020_polarisation}
Luo R.,  et~al., 2020, \mn@doi [Nature] {10.1038/s41586-020-2827-2}, 586, 693–696

\bibitem[\protect\citeauthoryear{{Lyutikov} \& {Popov}}{{Lyutikov} \& {Popov}}{2020}]{Lyutikov_2020}
{Lyutikov} M.,  {Popov} S.,  2020, \mn@doi [arXiv e-prints] {10.48550/arXiv.2005.05093}, \href {https://ui.adsabs.harvard.edu/abs/2020arXiv200505093L} {p. arXiv:2005.05093}

\bibitem[\protect\citeauthoryear{{MAGIC Collaboration} et~al.,}{{MAGIC Collaboration} et~al.}{2018}]{Magic_2018}
{MAGIC Collaboration} et~al., 2018, \mn@doi [Monthly Notices of the Royal Astronomical Society] {10.1093/mnras/sty2422}, 481, 2479

\bibitem[\protect\citeauthoryear{{Macquart} et~al.,}{{Macquart} et~al.}{2020}]{Macquart_2020}
{Macquart} J.~P.,  et~al., 2020, \mn@doi [\nat] {10.1038/s41586-020-2300-2}, \href {https://ui.adsabs.harvard.edu/abs/2020Natur.581..391M} {581, 391}

\bibitem[\protect\citeauthoryear{{Marcote} et~al.,}{{Marcote} et~al.}{2020}]{Marcote_2020}
{Marcote} B.,  et~al., 2020, \mn@doi [\nat] {10.1038/s41586-019-1866-z}, \href {https://ui.adsabs.harvard.edu/abs/2020Natur.577..190M} {577, 190}

\bibitem[\protect\citeauthoryear{{Margalit}, {Beniamini}, {Sridhar}  \& {Metzger}}{{Margalit} et~al.}{2020}]{Margalit_2020}
{Margalit} B.,  {Beniamini} P.,  {Sridhar} N.,   {Metzger} B.~D.,  2020, \mn@doi [\apjl] {10.3847/2041-8213/abac57}, \href {https://ui.adsabs.harvard.edu/abs/2020ApJ...899L..27M} {899, L27}

\bibitem[\protect\citeauthoryear{Metzger, Berger  \& Margalit}{Metzger et~al.}{2017}]{Metzger_2017}
Metzger B.~D.,  Berger E.,   Margalit B.,  2017, \mn@doi [The Astrophysical Journal] {10.3847/1538-4357/aa633d}, 841, 14

\bibitem[\protect\citeauthoryear{Metzger, Margalit  \& Sironi}{Metzger et~al.}{2019}]{Metzger_2019}
Metzger B.~D.,  Margalit B.,   Sironi L.,  2019, \mn@doi [Monthly Notices of the Royal Astronomical Society] {10.1093/mnras/stz700}, 485, 4091

\bibitem[\protect\citeauthoryear{Michilli et~al.,}{Michilli et~al.}{2018}]{Michilli_2018}
Michilli D.,  et~al., 2018, \mn@doi [Nature] {10.1038/nature25149}, 553, 182–185

\bibitem[\protect\citeauthoryear{{Mohan} \& {Rafferty}}{{Mohan} \& {Rafferty}}{2015}]{pybdsf}
{Mohan} N.,  {Rafferty} D.,  2015, {PyBDSF: Python Blob Detection and Source Finder}, Astrophysics Source Code Library, record ascl:1502.007

\bibitem[\protect\citeauthoryear{{Naletto} et~al.,}{{Naletto} et~al.}{2009}]{Naletto_2009}
{Naletto} G.,  et~al., 2009, \mn@doi [\aap] {10.1051/0004-6361/200912862}, \href {https://ui.adsabs.harvard.edu/abs/2009A&A...508..531N} {508, 531}

\bibitem[\protect\citeauthoryear{Newville et~al.,}{Newville et~al.}{2024}]{matt_newville_2024_12785036}
Newville M.,  et~al., 2024, lmfit/lmfit-py: 1.3.2, \mn@doi{10.5281/zenodo.12785036}, \url {https://doi.org/10.5281/zenodo.12785036}

\bibitem[\protect\citeauthoryear{{Niino} et~al.,}{{Niino} et~al.}{2022}]{Niino}
{Niino} Y.,  et~al., 2022, \mn@doi [\apj] {10.3847/1538-4357/ac6be8}, \href {https://ui.adsabs.harvard.edu/abs/2022ApJ...931..109N} {931, 109}

\bibitem[\protect\citeauthoryear{{Niu} et~al.,}{{Niu} et~al.}{2022}]{Niu_2022}
{Niu} C.~H.,  et~al., 2022, \mn@doi [\nat] {10.1038/s41586-022-04755-5}, \href {https://ui.adsabs.harvard.edu/abs/2022Natur.606..873N} {606, 873}

\bibitem[\protect\citeauthoryear{{N{\'u}{\~n}ez} et~al.,}{{N{\'u}{\~n}ez} et~al.}{2021}]{Nunez_2021}
{N{\'u}{\~n}ez} C.,  et~al., 2021, \mn@doi [\aap] {10.1051/0004-6361/202141110}, \href {https://ui.adsabs.harvard.edu/abs/2021A&A...653A.119N} {653, A119}

\bibitem[\protect\citeauthoryear{Offringa et~al.,}{Offringa et~al.}{2014}]{wsclean}
Offringa A.~R.,  et~al., 2014, \mn@doi [Monthly Notices of the Royal Astronomical Society] {10.1093/mnras/stu1368}, 444, 606

\bibitem[\protect\citeauthoryear{Pastor-Marazuela et~al.,}{Pastor-Marazuela et~al.}{2023}]{Pastor_Marazuela_2023}
Pastor-Marazuela I.,  et~al., 2023, \mn@doi [Astronomy &amp; Astrophysics] {10.1051/0004-6361/202243339}, 678, A149

\bibitem[\protect\citeauthoryear{{Pearlman} et~al.,}{{Pearlman} et~al.}{2023}]{Pearlman_2023}
{Pearlman} A.~B.,  et~al., 2023, \mn@doi [arXiv e-prints] {10.48550/arXiv.2308.10930}, \href {https://ui.adsabs.harvard.edu/abs/2023arXiv230810930P} {p. arXiv:2308.10930}

\bibitem[\protect\citeauthoryear{{Petroff} et~al.,}{{Petroff} et~al.}{2015}]{Petroff}
{Petroff} E.,  et~al., 2015, \mn@doi [\mnras] {10.1093/mnras/stu2419}, \href {https://ui.adsabs.harvard.edu/abs/2015MNRAS.447..246P} {447, 246}

\bibitem[\protect\citeauthoryear{{Planck Collaboration} et~al.,}{{Planck Collaboration} et~al.}{2020}]{Planck_2020}
{Planck Collaboration} et~al., 2020, \mn@doi [Astronomy &amp; Astrophysics] {10.1051/0004-6361/201833910}, 641, A6

\bibitem[\protect\citeauthoryear{Platts, Weltman, Walters, Tendulkar, Gordin  \& Kandhai}{Platts et~al.}{2019}]{Platts_2019}
Platts E.,  Weltman A.,  Walters A.,  Tendulkar S.,  Gordin J.,   Kandhai S.,  2019, \mn@doi [Physics Reports] {10.1016/j.physrep.2019.06.003}, 821, 1–27

\bibitem[\protect\citeauthoryear{Pleunis et~al.,}{Pleunis et~al.}{2021}]{Pleunis_2021}
Pleunis Z.,  et~al., 2021, \mn@doi [The Astrophysical Journal] {10.3847/1538-4357/ac33ac}, 923, 1

\bibitem[\protect\citeauthoryear{{Prochaska} \& {Zheng}}{{Prochaska} \& {Zheng}}{2019}]{Prochaska_2019}
{Prochaska} J.~X.,  {Zheng} Y.,  2019, \mn@doi [\mnras] {10.1093/mnras/stz261}, \href {https://ui.adsabs.harvard.edu/abs/2019MNRAS.485..648P} {485, 648}

\bibitem[\protect\citeauthoryear{Prochaska et~al.,}{Prochaska et~al.}{2023}]{j_xavier_prochaska_2023_8125230}
Prochaska J.~X.,  et~al., 2023, FRBs/FRB: Release to sync with Gordon et al. 2023, \mn@doi{10.5281/zenodo.8125230}, \url {https://doi.org/10.5281/zenodo.8125230}

\bibitem[\protect\citeauthoryear{{Purcell}, {Van Eck}, {West}, {Sun}  \& {Gaensler}}{{Purcell} et~al.}{2020}]{RM_tools}
{Purcell} C.~R.,  {Van Eck} C.~L.,  {West} J.,  {Sun} X.~H.,   {Gaensler} B.~M.,  2020, {RM-Tools: Rotation measure (RM) synthesis and Stokes QU-fitting}, Astrophysics Source Code Library, record ascl:2005.003

\bibitem[\protect\citeauthoryear{Rajwade et~al.,}{Rajwade et~al.}{2022}]{Rajwade_2022}
Rajwade K.~M.,  et~al., 2022, \mn@doi [Monthly Notices of the Royal Astronomical Society] {10.1093/mnras/stac1450}, 514, 1961–1974

\bibitem[\protect\citeauthoryear{{Rajwade} et~al.,}{{Rajwade} et~al.}{2024}]{Rajwade_2024}
{Rajwade} K.~M.,  et~al., 2024, \mn@doi [arXiv e-prints] {10.48550/arXiv.2407.02173}, \href {https://ui.adsabs.harvard.edu/abs/2024arXiv240702173R} {p. arXiv:2407.02173}

\bibitem[\protect\citeauthoryear{{Ravi} et~al.,}{{Ravi} et~al.}{2023}]{Ravi_2023}
{Ravi} V.,  et~al., 2023, \mn@doi [\apjl] {10.3847/2041-8213/acc4b6}, \href {https://ui.adsabs.harvard.edu/abs/2023ApJ...949L...3R} {949, L3}

\bibitem[\protect\citeauthoryear{{Ryder} et~al.,}{{Ryder} et~al.}{2023}]{Ryder}
{Ryder} S.~D.,  et~al., 2023, \mn@doi [Science] {10.1126/science.adf2678}, \href {https://ui.adsabs.harvard.edu/abs/2023Sci...382..294R} {382, 294}

\bibitem[\protect\citeauthoryear{{Sako} et~al.,}{{Sako} et~al.}{2018}]{Tomo-e}
{Sako} S.,  et~al., 2018, in {Evans} C.~J.,  {Simard} L.,   {Takami} H.,  eds,  Society of Photo-Optical Instrumentation Engineers (SPIE) Conference Series Vol. 10702, Ground-based and Airborne Instrumentation for Astronomy VII. p. 107020J, \mn@doi{10.1117/12.2310049}

\bibitem[\protect\citeauthoryear{Sanidas, Caleb, Driessen, Morello, Rajwade  \& Stappers}{Sanidas et~al.}{2017}]{sanidas_2018}
Sanidas S.,  Caleb M.,  Driessen L.,  Morello V.,  Rajwade K.,   Stappers B.,  2017, \mn@doi [Proceedings of the International Astronomical Union] {10.1017/S1743921317009310}, 13, 406

\bibitem[\protect\citeauthoryear{{Schlafly} \& {Finkbeiner}}{{Schlafly} \& {Finkbeiner}}{2011}]{Schlafly}
{Schlafly} E.~F.,  {Finkbeiner} D.~P.,  2011, \mn@doi [\apj] {10.1088/0004-637X/737/2/103}, \href {https://ui.adsabs.harvard.edu/abs/2011ApJ...737..103S} {737, 103}

\bibitem[\protect\citeauthoryear{{Schwab}}{{Schwab}}{1984}]{Schwab_1984}
{Schwab} F.~R.,  1984, \mn@doi [\aj] {10.1086/113605}, \href {https://ui.adsabs.harvard.edu/abs/1984AJ.....89.1076S} {89, 1076}

\bibitem[\protect\citeauthoryear{{Scott}}{{Scott}}{2018}]{Scott_2018}
{Scott} N.~J.,  2018, in American Astronomical Society Meeting Abstracts \#231. p. 361.16

\bibitem[\protect\citeauthoryear{{Scott} et~al.,}{{Scott} et~al.}{2021}]{Scott_2021}
{Scott} N.~J.,  et~al., 2021, \mn@doi [Frontiers in Astronomy and Space Sciences] {10.3389/fspas.2021.716560}, \href {https://ui.adsabs.harvard.edu/abs/2021FrASS...8..138S} {8, 138}

\bibitem[\protect\citeauthoryear{{Shannon} et~al.,}{{Shannon} et~al.}{2024}]{shannon2024}
{Shannon} R.~M.,  et~al., 2024, \mn@doi [arXiv e-prints] {10.48550/arXiv.2408.02083}, \href {https://ui.adsabs.harvard.edu/abs/2024arXiv240802083S} {p. arXiv:2408.02083}

\bibitem[\protect\citeauthoryear{{Spitler} et~al.,}{{Spitler} et~al.}{2016}]{Spitler_2016}
{Spitler} L.~G.,  et~al., 2016, \mn@doi [\nat] {10.1038/nature17168}, \href {https://ui.adsabs.harvard.edu/abs/2016Natur.531..202S} {531, 202}

\bibitem[\protect\citeauthoryear{Szentgyorgyi et~al.,}{Szentgyorgyi et~al.}{2005}]{szentgyorgyi2005keplercam}
Szentgyorgyi A.,  et~al., 2005, in American Astronomical Society Meeting Abstracts. pp 110--10

\bibitem[\protect\citeauthoryear{{Tendulkar} et~al.,}{{Tendulkar} et~al.}{2017}]{Tendulkar_2017}
{Tendulkar} S.~P.,  et~al., 2017, \mn@doi [\apjl] {10.3847/2041-8213/834/2/L7}, \href {https://ui.adsabs.harvard.edu/abs/2017ApJ...834L...7T} {834, L7}

\bibitem[\protect\citeauthoryear{{The Dark Energy Survey Collaboration}}{{The Dark Energy Survey Collaboration}}{2005}]{thedarkenergysurveycollaboration2005dark}
{The Dark Energy Survey Collaboration} 2005, The Dark Energy Survey (\mn@eprint {arXiv} {astro-ph/0510346})

\bibitem[\protect\citeauthoryear{Thornton et~al.,}{Thornton et~al.}{2013}]{Thornton_2013}
Thornton D.,  et~al., 2013, \mn@doi [Science] {10.1126/science.1236789}, 341, 53–56

\bibitem[\protect\citeauthoryear{{Trudu} et~al.,}{{Trudu} et~al.}{2023}]{trudu}
{Trudu} M.,  et~al., 2023, \mn@doi [\aap] {10.1051/0004-6361/202245303}, \href {https://ui.adsabs.harvard.edu/abs/2023A&A...676A..17T} {676, A17}

\bibitem[\protect\citeauthoryear{{Vreeswijk} \& {Paterson}}{{Vreeswijk} \& {Paterson}}{2021a}]{blackbox}
{Vreeswijk} P.,  {Paterson} K.,  2021a, {BlackBOX: BlackGEM and MeerLICHT image reduction software}, Astrophysics Source Code Library, record ascl:2105.011

\bibitem[\protect\citeauthoryear{{Vreeswijk} \& {Paterson}}{{Vreeswijk} \& {Paterson}}{2021b}]{zogy}
{Vreeswijk} P.,  {Paterson} K.,  2021b, {ZOGY: Python implementation of proper image subtraction}, Astrophysics Source Code Library, record ascl:2105.010

\bibitem[\protect\citeauthoryear{Wadiasingh \& Timokhin}{Wadiasingh \& Timokhin}{2019}]{Wadiasingh_2019}
Wadiasingh Z.,  Timokhin A.,  2019, \mn@doi [The Astrophysical Journal] {10.3847/1538-4357/ab2240}, 879, 4

\bibitem[\protect\citeauthoryear{{Xing} \& {Yu}}{{Xing} \& {Yu}}{2024}]{gamma-ray}
{Xing} Y.,  {Yu} W.,  2024, The Astronomer's Telegram, \href {https://ui.adsabs.harvard.edu/abs/2024ATel16630....1X} {16630, 1}

\bibitem[\protect\citeauthoryear{Yamasaki \& Totani}{Yamasaki \& Totani}{2020}]{Yamasaki_2020}
Yamasaki S.,  Totani T.,  2020, \mn@doi [The Astrophysical Journal] {10.3847/1538-4357/ab58c4}, 888, 105

\bibitem[\protect\citeauthoryear{Yang \& Zhang}{Yang \& Zhang}{2018}]{Yang_2018}
Yang Y.-P.,  Zhang B.,  2018, \mn@doi [The Astrophysical Journal] {10.3847/1538-4357/aae685}, 868, 31

\bibitem[\protect\citeauthoryear{Yao, Manchester  \& Wang}{Yao et~al.}{2017}]{Yao_2017}
Yao J.~M.,  Manchester R.~N.,   Wang N.,  2017, \mn@doi [The Astrophysical Journal] {10.3847/1538-4357/835/1/29}, 835, 29

\bibitem[\protect\citeauthoryear{{York} et~al.,}{{York} et~al.}{2000}]{SDSS}
{York} D.~G.,  et~al., 2000, \mn@doi [\aj] {10.1086/301513}, \href {https://ui.adsabs.harvard.edu/abs/2000AJ....120.1579Y} {120, 1579}

\bibitem[\protect\citeauthoryear{Zackay, Ofek  \& Gal-Yam}{Zackay et~al.}{2016}]{Zackay_2016}
Zackay B.,  Ofek E.~O.,   Gal-Yam A.,  2016, \mn@doi [The Astrophysical Journal] {10.3847/0004-637x/830/1/27}, 830, 27

\bibitem[\protect\citeauthoryear{{Zampieri} et~al.,}{{Zampieri} et~al.}{2015}]{Zampieri_2015}
{Zampieri} L.,  et~al., 2015, in {Prochazka} I.,  {Sobolewski} R.,   {James} R.~B.,  eds,  Society of Photo-Optical Instrumentation Engineers (SPIE) Conference Series Vol. 9504, Photon Counting Applications 2015. p. 95040C (\mn@eprint {arXiv} {1505.07339}), \mn@doi{10.1117/12.2179547}

\bibitem[\protect\citeauthoryear{{Zhang}}{{Zhang}}{2018}]{ZhangBing_2018}
{Zhang} B.,  2018, \mn@doi [\apjl] {10.3847/2041-8213/aae8e3}, \href {https://ui.adsabs.harvard.edu/abs/2018ApJ...867L..21Z} {867, L21}

\bibitem[\protect\citeauthoryear{Zhang}{Zhang}{2023}]{Zhang_2023}
Zhang B.,  2023, \mn@doi [Reviews of Modern Physics] {10.1103/revmodphys.95.035005}, 95

\bibitem[\protect\citeauthoryear{{Zhang} \& {Zhang}}{{Zhang} \& {Zhang}}{2017}]{Zhang}
{Zhang} B.-B.,  {Zhang} B.,  2017, \mn@doi [\apjl] {10.3847/2041-8213/aa7633}, \href {https://ui.adsabs.harvard.edu/abs/2017ApJ...843L..13Z} {843, L13}

\bibitem[\protect\citeauthoryear{Zou et~al.,}{Zou et~al.}{2022}]{Zou_2022}
Zou H.,  et~al., 2022, \mn@doi [Research in Astronomy and Astrophysics] {10.1088/1674-4527/ac6416}, 22, 065001

\makeatother
\end{thebibliography}

\bsp
\label{lastpage}
\end{document}